\documentstyle[epsf]{mn}
\begin{document}

\title[Broadband power spectra of AGN]
{Measuring the broadband power spectra of active galactic nuclei with RXTE}
\author[P. Uttley, I. M. M$^{\rm c}$Hardy and I. E. Papadakis]
{P. Uttley$^{1}$\thanks{e-mail: pu@astro.soton.ac.uk}, 
I. M. M$^{\rm c}$Hardy$^{1}$ and I. E. Papadakis$^{2}$ \\
$^{1}$Department of Physics and Astronomy, University of Southampton, 
Southampton SO17 1BJ \\
$^{2}$Physics Department, University of Crete, PO Box 2208, 710 03
Heraklion, Crete, Greece \\
}

\date{}

\maketitle
\parindent 18pt
\begin{abstract} 
We have developed a Monte Carlo technique to
test models for the true power spectra of intermittently
sampled lightcurves against the noisy,
observed power spectra, and produce a reliable
estimate of the goodness of fit of the given model.  We apply this technique
to constrain the broadband power spectra of a sample of four Seyfert 
galaxies monitored by the {\it Rossi X-ray Timing Explorer} ({\it RXTE})
over three years.  We show that the power spectra of three of the AGN 
in our sample (MCG-6-30-15, NGC~5506 and NGC~3516) flatten
significantly towards low frequencies, while the power spectrum of 
NGC~5548 shows no evidence of flattening.  We fit two models for the
flattening, a `knee' model, analogous to the low-frequency break seen in
the power spectra of BHXRBs in the low state
(where the power-spectral slope flattens to $\alpha=0$)
 and a `high-frequency break' model (where the power-spectral
slope flattens to $\alpha=1$), analogous to 
the high-frequency break seen in the high and low-state 
power spectra of the classic BHXRB Cyg~X-1. 
Both models provide good fits to the power spectra of all four AGN. 
For both models, the
characteristic frequency for flattening is significantly higher in
MCG-6-30-15 than in NGC~3516 (by factor $\sim10$) although both
sources have similar X-ray luminosities, suggesting that MCG-6-30-15 has
a lower black hole mass and is accreting at a higher rate than NGC~3516.
Assuming linear scaling of characteristic frequencies with black hole
mass, the high accretion rate implied for MCG-6-30-15 favours the
high-frequency break model for this source and further suggests that 
MCG-6-30-15 and possibly NGC~5506, may be analogues of Cyg~X-1 in the 
high state.  Comparison
of our model fits with naive fits, where the model is fitted 
directly to the observed power spectra (with errors estimated from the 
data), shows that Monte Carlo fitting is essential for reliably
constraining the broadband power spectra of AGN lightcurves obtained to date.  
\end{abstract}


\section{Introduction}
\label{intro}
The strong and rapid X-ray variability observed in many Seyfert galaxies on
time-scales of a day or less provides strong evidence that the X-rays
are emitted close to the central black hole.  Early efforts to
characterise the X-ray variability, using data from {\it EXOSAT}, showed
that it has a scale-invariant, red-noise form on
time-scales from a few hundred seconds up to the few-days duration of the
observations (M$^{\rm c}$Hardy
\& Czerny 1987, Lawrence et al. 1987).  Later studies of the X-ray variability
properties of large samples of radio-quiet AGN showed that the
variability amplitude scales inversely with luminosity (Green, M$^{\rm
c}$Hardy \& Lehto 1993; Lawrence \& Papadakis 1993; Nandra et al. 1997).
 One possible explanation of this result is that the higher luminosity
AGN contain more massive black holes and the 
variability time-scales in AGN scale with black hole mass.
Intriguingly, black hole X-ray binary
systems (BHXRBs) also show red-noise type variability of a similar
amplitude to AGN, on time-scales less than seconds.  The similarity in
X-ray variability properties of AGN and BHXRBs raises the
possibility that the processes causing variability in AGN and BHXRBs are
the same and that any characteristic variability time-scales scale with
the central black hole mass. 
This possibility can be tested by comparing the detailed X-ray timing
properties of BHXRBs and AGN. \\
Timing studies of BHXRBs are usually carried out in the frequency
domain using the power spectrum, which shows the contribution
of variations on different time-scales (corresponding to 
power-spectral frequencies) to the total variability of the lightcurve. 
The power spectra of BHXRBs are dominated by a broadband noise
component (van der Klis 1995).  On short time-scales, the variability is characterised as
scale-invariant `red noise', producing a power-law power spectrum (power
$P(\nu)$ at frequency $\nu$ is given by
$P(\nu)\propto \nu^{-\alpha}$ where $\alpha$ is the power-spectral
slope) of slope $\alpha \sim1$--$2.5$ (van der Klis 1995). 
In the `low' state, characterised by a relatively hard X-ray 
spectrum (similar to that of AGN), the power
spectrum flattens towards lower frequencies so that on long time-scales
the X-ray lightcurve becomes `white noise', with corresponding slope
$\alpha \simeq0$. 
For example, in the classic BHXRB system Cyg~X-1, the power-spectral
flattening is well described by a power-law
with two breaks, a high-frequency break which varies between 1
and 6~Hz, above which the power-spectral slope varies between $\alpha
\sim1.5$--$2.4$ and a low-frequency break
which varies between 0.04 and 0.4~Hz, above which the power-spectral
slope $\alpha \simeq1$ and below which the slope $\alpha \simeq0$
(Belloni \& Hasinger 1990).  In contrast, the power spectrum of
the `high' (soft energy spectrum) state seen in some BHXRBs (inluding
Cyg~X-1) does not flatten to zero slope;
instead, the slope $\alpha\simeq1$ below the high-frequency break extends to
$<10^{-2}$~Hz (e.g. Cui et al. 1997). \\
In order to test the hypothesis that the X-ray variability of AGN
is similar to that of BHXRBs over a broad range of time-scales, we
must search for low-frequency flattening in the broadband power spectra of AGN.
By fitting models with power-spectral breaks to the AGN power spectra and
comparing the estimated break frequencies with what we expect if they
correspond to similar breaks in BHXRB power spectra, we can test the
possibility that the power-spectral shape is really the same and scales
simply with black hole mass.
If so, we expect break frequencies in AGN to be found at
frequencies of $\sim10^{-5}$~Hz or lower, so that monitoring
observations on time-scales of weeks or longer are necessary to detect
any flattening in the power spectrum. \\
Early attempts to measure broadband power spectra of AGN were hampered
by the sparseness of long-term archival lightcurves, which had to be
constructed from data obtained by several missions (M$^{\rm c}$Hardy 1988). 
Nonetheless, some
evidence for power-spectral flattening was found, but models for the form
of the flattening could not be constrained (Papadakis \& M$^{\rm
c}$Hardy 1995). \\
Ideally, broadband power
spectra should be measured from lightcurves obtained with frequent and regular
sampling over a long duration, which previous missions were not
optimised to do.  The {\it Rossi X-ray Timing
Explorer} ({\it RXTE}), which has just such a capability, was launched
in December 1995.  {\it RXTE} carries a large-area proportional counter
array (the PCA) which can detect many AGN with good signal-to-noise in less
than 1000~s, but most importantly, {\it RXTE} can slew rapidly
so that it may monitor many targets with frequent 1~ks snapshots. \\
We have monitored a sample of 4 Seyfert galaxies (MCG-6-30-15,
NGC~4051, 5506 and 5548) with {\it RXTE} since 1996, 
in order to measure their broadband power spectra.  These objects are
known to be significantly X-ray variable and
cover a broad range of X-ray luminosity (NGC~4051$\sim5\times 10^{41}$~erg~s$^{-1}$,
MCG-6-30-15 and NGC~5506$\sim1.5\times 10^{43}$~erg~s$^{-1}$,
NGC~5548$\sim5\times10^{43}$~erg~s$^{-1}$) and presumably, a broad range
of black hole masses.  We describe the
power spectrum of NGC~4051, which shows unusual non-stationarity in its
lightcurve (Uttley et al. 1999) in a separate paper (Papadakis, M$^{\rm
c}$Hardy \& Uttley, in prep.).  In this paper, we present a
power-spectral study of the remaining three objects in our sample using
data from {\it RXTE} cycles 1, 2 and 3, also including
the excellent lightcurves obtained as part of a separate power-spectral 
study of the Seyfert~1 galaxy NGC~3516
(luminosity$\sim1.5\times 10^{43}$~erg~s$^{-1}$), by Edelson \& Nandra (1999).
 We describe data reduction and present the lightcurves in Section~2. \\
The estimation of the
underlying power-spectral shape from lightcurves which are
discretely (and possibly unevenly) sampled is hampered by the distorting
effects of aliasing and red-noise leak.  A further serious problem is
that the measured power spectra are intrinsically noisy, and
reliable errors on the power in each frequency bin
cannot be estimated from the data (especially at low frequencies),
due to the small number ($<20$) of power-spectral measurements made in
each frequency bin.  In Section~3, after presenting the observed power spectra, 
we describe these problems, which previous efforts to constrain
the shape of the broadband power spectrum of AGN using {\it RXTE} data 
(e.g. Edelson \& Nandra 1999, Nowak \& Chiang 2000) have not accounted for. \\
To overcome the difficulties in estimating the true power-spectral shape,
we have developed a method which we call {\sc psresp}, based on the 
response method (Done et al. 1992) which uses Monte Carlo simulations of
lightcurves to take account of the distorting effects of sampling and
to estimate uncertainties, allowing us to test various power-spectral
models against the data.  We describe {\sc psresp} in Section~4, and
apply it to the lightcurves of our sample of
Seyfert galaxies in Section~5, in order to test for flattening in their
broadband power spectra and constrain simple models for describing any
flattening we see.  In Section~6, we compare our results with those
obtained by naively fitting the observed power spectrum (without taking
proper account of errors and the distortion due to sampling), use our power
spectral measurements to estimate the black hole masses of the AGN in our
sample and discuss some of the implications of our results, before making
concluding remarks in Section~7.

\section{The lightcurves}
\subsection{Observations and data reduction}
\begin{figure*}
\begin{center}
{\epsfxsize 0.7\hsize
 \leavevmode
 \epsffile{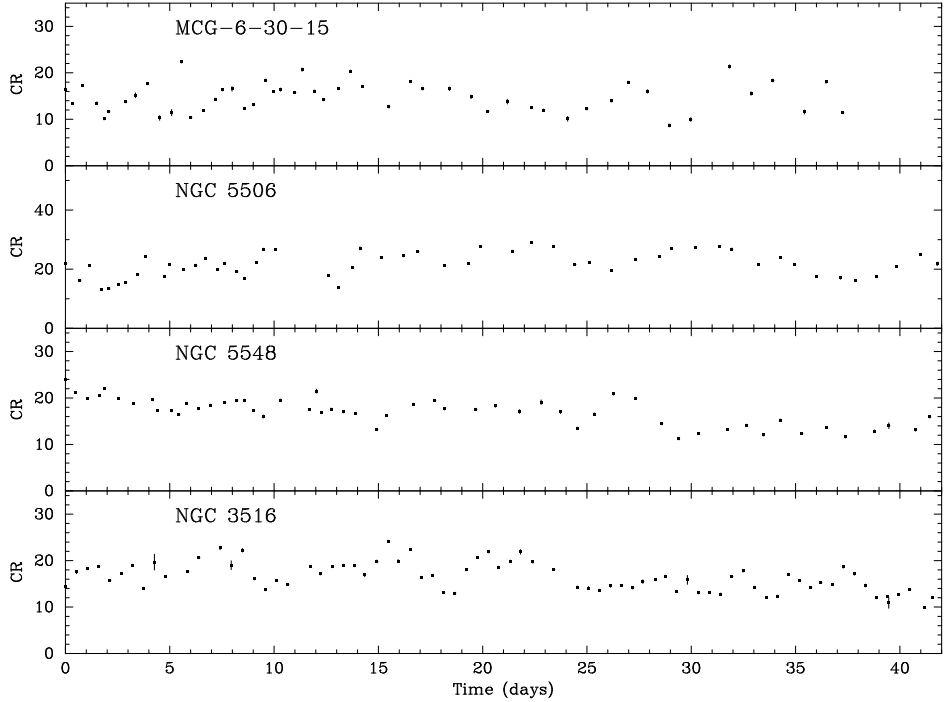}
}\caption{2--10~keV intensive monitoring lightcurves of MCG-6-30-15, NGC~5506,
NGC~5548 and NGC~3516.}
\label{fig:day}
\end{center}
\end{figure*}

\begin{figure*}
\begin{center}
{\epsfxsize 0.7\hsize
 \leavevmode
 \epsffile{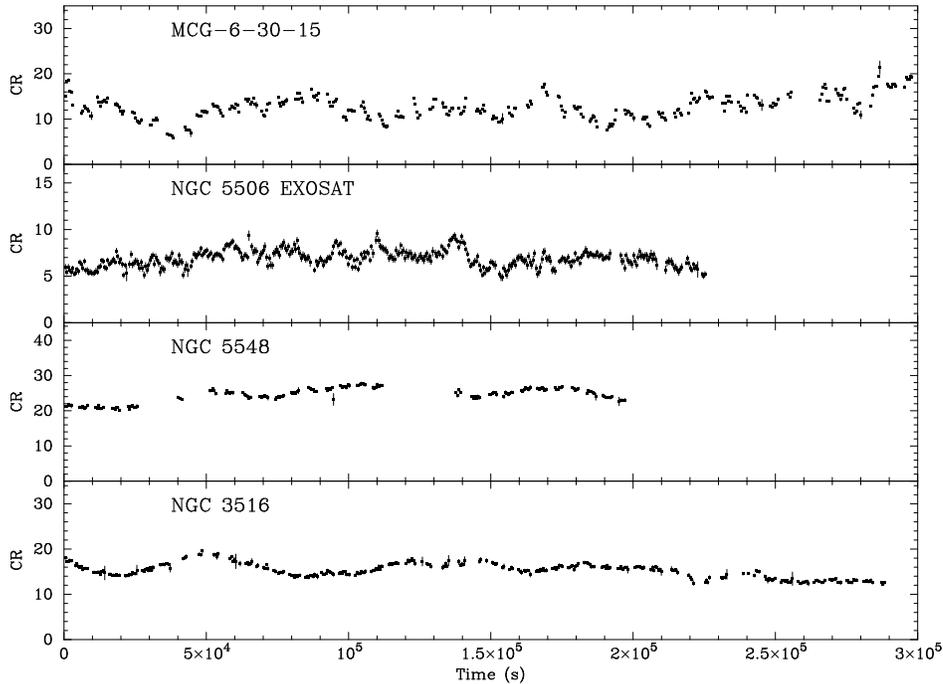}
}\caption{2--10~keV long-look lightcurves of MCG-6-30-15, NGC~5506 ({\it
EXOSAT} 1--9~keV),
NGC~5548 and NGC~3516.}
\label{fig:long}
\end{center}
\end{figure*}

We use data obtained with the PCA on board {\it RXTE} covering a three
year period during observing cycles
1--3 when the PCA gain setting was constant, so that count rate
measurements provide a simple measure of observed flux. 
Of the three instruments on board {\it RXTE}, only the PCA is
sensitive enough to allow us to make an accurate flux measurement
for our targets with the 1~ks snapshots which our monitoring consists
of.  The PCA consists of 5 Xenon-filled Proportional Counter Units
(PCUs), numbered 0 to 4 which are sensitive in the 2--60~keV energy
range and contribute to a total effective area of 6500~cm$^{2}$.  Since
launch, discharge problems have meant that one or both of PCUs 3 and 4
are often switched off (this problem extended to PCU~1 in March
1999 but we do not include this later data here).  Despite the loss of up
to two PCUs during our observations, we
are easily able to obtain sufficient signal-to-noise in a single snapshot
for our purposes ($S/N>20\sigma$). \\
\begin{table*}
\caption{Lightcurve details}
\label{tab:lcs}
\begin{tabular}{lccccccc}
 & \multicolumn{2}{c}{Long-term} & \multicolumn{2}{c}{Intensive} &
\multicolumn{3}{c}{Long-look} \\
 & Start & Stop & Start & Stop & Start & Stop & Exp \\
MCG-6-30-15 & 8 May 1996 & 2 Feb 1999 & 23 Aug 1996 & 29 Sep 1996 &
03:31 {\sc ut} 4 Aug 1997 & 12:34 {\sc ut} 12 Aug 1997 & 332 \\
NGC~5506 & 23 Apr 1996 & 2 Feb 1999 & 8 Aug 1996 & 19 Sep 1996 &
04:45 {\sc ut} 20 Jun 1997 & 12:33 {\sc ut} 9 Jul 1997 & 93 \\
NGC~5548 & 23 Apr 1996 & 22 Dec 1998 & 26 Jun 1996 & 8 Aug 1996 & 
12:41 {\sc ut} 19 Jun 1998 & 07:01 {\sc ut} 24 Jun 1998 & 99 \\  
NGC~3516 & 16 Mar 1997 & 28 Dec 1998 & 16 Mar 1997 & 30 Jul 1997 &
00:14 {\sc ut} 22 May 1997 & 05:37 {\sc ut} 26 May 1997 & 249 \\
 \end{tabular}

\medskip
\raggedright{The table shows the start and stop times of the lightcurves
used in this work (except 2nd NGC~3516 long-look - see text for details. 
Also given is the useful exposure
time in ks (Exp) for each long-look observation.}
\end{table*}
In order to efficiently measure a power spectrum over the broadest
range of time-scales while minimising the necessary observing time, we
monitored our targets using several different schemes, each designed to
measure the power spectrum over a different frequency range.  In
1996, we observed MCG-6-30-15, NGC~5506 and NGC~5548 twice daily for
$\sim2$~weeks followed by daily observations for $\sim4$~weeks and then
weekly for the remainder of the year.  During the following two years,
we observed our targets every two weeks.  NGC~3516 was monitored as part
of a separate study with broadly similar goals to our own (Edelson \& Nandra
1999).  Here we use public archival data from this campaign, including
an intensive period of monitoring every 12.8~h for $\sim$4~months
duration, and long-term monitoring at 4.3~d intervals from March 1997 until
the end of 1998.  The start and end dates of all the lightcurves
are shown in Table~\ref{tab:lcs} \\
We measure variability on short time-scales using `long-look'
observations, quasi-continuous observations of duration $\sim$days,
which we obtained ourselves (NGC~5506) or from the {\it RXTE} public
archive (MCG-6-30-15, NGC~5548 and NGC~3516). 
The details of these observations are also summarised in Table~\ref{tab:lcs}. 
Not shown in Table~\ref{tab:lcs} are details of a second long-look
observation of NGC~3516, obtained from
08:00~{\sc ut} 13 April to 16:13~{\sc ut} 16 April 1998 (148~ks useful
exposure), which we also
include to maximise the definition of the power spectrum of NGC~3516
at high frequencies.
Unfortunately, the NGC~5506 long-look is too sparsely 
sampled (spread over a 20 day period) to be useful for
measuring the power spectrum except at the highest frequencies
($\sim10^{-3}$~Hz). 
Therefore, in order to measure the power spectrum of NGC~5506 in the
$10^{-5}$--$10^{-3}$~Hz range, we use an archival {\it EXOSAT} ME 
lightcurve, of $\sim230$~ks continuous duration, obtained during 24-27 January 1986
and originally described by M$^{\rm c}$Hardy \& Czerny (1987).  The
energy range sampled by {\it EXOSAT} (1--9~keV) is comparable to the
2--10~keV range which we will measure with {\it RXTE}, so that the
normalised power spectrum should have a similar shape
and amplitude to that measured by {\it RXTE}, if the power-spectral
shape is stationary on time-scales of a decade (see
Section~\ref{station}). \\
We reduce all {\it RXTE} data using {\sc ftools v4.2}.  Because PCUs 3
and 4 are often switched off, we only use data from PCUs 0, 1 and 2.  We
extract data from the top layer only (to minimise background relative to
source counts) and make lightcurves in the 2--10~keV channel range
corresponding to absolute channels 7--28.  We exclude data obtained
within and up to 20 minutes after SAA maximum and data obtained with
earth elevation $<10^{\circ}$, target offset $>0.02^{\circ}$
and electron contamination $>10\%$.  We estimate background lightcurves
using the L7 background model. \\
\begin{figure*}
\begin{center}
{\epsfxsize 0.7\hsize
 \leavevmode
 \epsffile{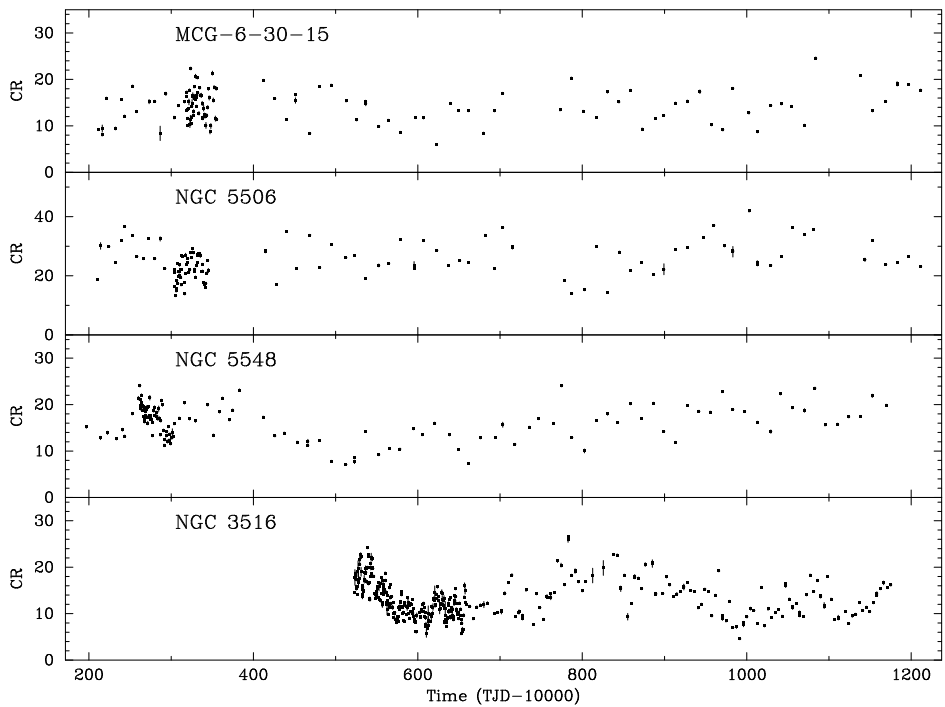}
}\caption{2--10~keV monitoring lightcurves of MCG-6-30-15, NGC~5506,
NGC~5548 and NGC~3516.  CR units are count~s$^{-1}$.} \label{fig:mon}
\end{center}
\end{figure*}
We show the long-term monitoring lightcurves in Figure~\ref{fig:mon}. 
The annual gaps lasting $\sim6$--8~weeks in the MCG-6-30-15 and NGC~5506
lightcurves correspond to periods when sun-angle constraints prevent
{\it RXTE} from pointing at these objects. 
Strong variability can be seen in all four lightcurves, and long-term
trends are particularly apparent in the lightcurves of NGC~5548 and
NGC~3516.  \\
It is important to note that the quality of these long-term
monitoring lightcurves is far superior to that obtainable with the
All-sky monitor (ASM) on board {\it RXTE} which, although excellent for
monitoring bright sources, is subject to large systematic errors when
observing faint sources like the AGN we study here.  This is apparent if
we compare the 28-day averaged ASM and PCA lightcurves of NGC~5548 
obtained over the same period (see Fig.~\ref{fig:asmpca}). 
The ASM lightcurve looks very different
to the PCA lightcurve, therefore ASM data should not be used to measure
the low-frequency power spectra of faint sources ($<0.5$~ASM~
count~s$^{-1}$). \\
\begin{figure}
\begin{center}
{\epsfxsize 0.9\hsize
 \leavevmode
 \epsffile{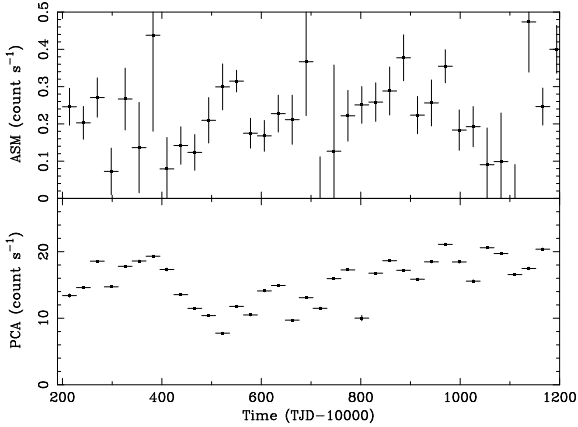}
}\caption{Comparison of ASM and PCA lightcurves of NGC~5548.  The
lightcurves are averaged in 28~d bins.  ASM error bars represent
standard errors.  No errors are determined for the PCA lightcurve, although the
small spread of points on short time-scales in the original lightcurve
(Fig.~\ref{fig:mon}) suggests that the uncertainties are small.} \label{fig:asmpca}
\end{center}
\end{figure}
A close-up look at the period of intensive (twice-daily and
daily) PCA monitoring can be seen in Figure~\ref{fig:day}, which is plotted
in terms of days since the start of each intensive monitoring period. 
The NGC~3516 intensive monitoring lightcurve is cut short so that
the lightcurves are of similar length for comparison purposes (see
Edelson \& Nandra 1999, for the full lightcurve).  Occasional short gaps in the
lightcurves are due to observations excluded due to our data extraction
criteria.  Significant variability on
time-scales of days can be seen in all four lightcurves, but
MCG-6-30-15 shows the strongest variations on the shortest time-scales.\\
We show the long-look lightcurves in Figure~\ref{fig:long},
binned to 512~s resolution.
For comparison purposes, we plot similar lengths of lightcurves and
cut off more than half of the MCG-6-30-15 lightcurve
(which can be seen in full in Lee et al. 1999).  We plot only the April
1998 long-look observation of NGC~3516 (see Edelson \& Nandra 1999 for
the earlier long-look lightcurve).  We show here the continuous {\it EXOSAT} 
lightcurve of NGC~5506 for comparison purposes (see Lamer, Uttley \&
M$^{\rm c}$Hardy 2000 for the {\it RXTE} lightcurve).  On short time-scales,
it can be seen that the MCG-6-30-15 and NGC~5506 lightcurves look
similar and show quite strong, rapid variability.  On the other hand, NGC~5548
and NGC~3516 show slower, more gradual trends. 

\subsection{Background subtraction and source contamination}
Because the PCA is not an imaging instrument, the contribution of
background to the lightcurves must be modelled.  Discrepancies between
the model background and the real background might then contaminate the
background-subtracted lightcurves.  A further source of contamination
may be due to other, reasonably bright sources in the field of view.  We now
briefly consider the possible contribution of this contamination
to our lightcurves. \\
Edelson \& Nandra (1999) use offset
pointings to show that the average discrepancy between the L7 background model
and the measured background in the 2--10~keV band is significant (0.87~count~s$^{-1}$) but
varies little (noise subtracted RMS 0.39~count~s$^{-1}$) and so
introduces little power into the measured power spectrum.  Moreover, the
variations in this background error occur only on long time-scales
(weeks) where the source variability is stronger, so we do not expect
any spurious power introduced by these variations to be significant
compared to the power intrinsic to the source.  However, Uttley et al. (1999)
show that spectra of NGC~4051, obtained simultaneously while it was very faint
($1.3\times10^{-12}$~erg~cm$^{-2}$~s$^{-1}$) by {\it RXTE} and the
imaging MECS intruments on board {\it BeppoSAX}
are in good agreement with one another, implying little
background offset (since the total 2--10~keV source count rate observed by {\it RXTE}
in 3 PCUs was only 0.4~count~s$^{-1}$).  This discrepancy with the
significant background offset observed in NGC~3516 implies that the offset
is dependent on the source being observed, and hence may be
associated with spatial fluctuations in the cosmic X-ray background or other
faint sources in the {\it RXTE} field of view.  Any small constant offset due
to inaccurate background modelling will only affect the normalisation of
the power spectrum by a relatively small amount 
(once it has been normalised by squared mean flux, see
Section~\ref{powmeas}) and will not affect the shape of the power
spectrum at all, so we do not consider it further in the cases of
MCG-6-30-15, NGC~5506 and NGC~3516.  We note however that the
observations of NGC~5548 suffer a minor complication, in that the field
of view also contains the bright BL~Lac object 1E~1415.6+2557, offset
0.5$^{\circ}$ from NGC~5548.  Chiang et al. (2000) conducted separate
pointings at this source and found that its
contaminating contribution to the measured 2--10~keV PCA count rate
(for 3 PCUs) of NGC~5548, after allowing for the effects of the PCA 
collimators, was only $\sim2$~count~s$^{-1}$ (about 10\% of the total
measured count rate).  The contaminating flux was estimated to vary by
$\leq0.8$~count~s$^{-1}$ in two months so that, assuming that there is not
much stronger variability on longer time-scales, 1E~1415.6+2557
should not contribute significantly to the low-frequency power measured from
the {\it RXTE} lightcurve.  However,
in order to take account of the contaminating contribution to the mean flux
level of the NGC~5548 lightcurve, we shall subtract
2~count~s$^{-1}$ from the measured 2--10~keV mean flux level of NGC~5548
for the purposes of power-spectral normalisation.

\section{The power spectra}
\label{powiss}
\subsection{Measuring the raw power spectra}
\label{powmeas}
To obtain the power spectrum of a discretely and possibly unevenly sampled 
light curve $f(t_{i})$, of length $N$ data points, we first subtract the
mean flux $\mu$ from the lightcurve (to remove zero-frequency power) and then
calculate the modulus squared of its discrete Fourier transform at
each sampled frequency $\nu$ (e.g. Deeming 1975):
\[
|F_{N}(\nu )|^{2}=\! \left(\sum_{i=1}^{N}f(t_{i})\,\cos(2 \, \pi \, \nu \,
t_{i})\right)^{2}\!+\!\left(\sum_{i=1}^{N}f(t_{i})\,\sin(2 \, \pi \, \nu \,
t_{i})\right)^{2}.
\]
Note that the frequencies sampled by the discrete Fourier transform
occur at evenly spaced intervals, $\nu_{\rm min}, 2\nu_{\rm
min}, 3\nu_{\rm min},....\nu_{\rm Nyq}$, where $\nu_{\rm min}$ is equal
to $T^{-1}$ (where $T$ is the total duration of the lightcurve,
i.e. $T=t_{N}-t_{1}$) and the Nyquist frequency $\nu_{\rm Nyq}=(2T/N)^{-1}$.
 We obtain the power $P(\nu)$ by applying a suitable normalisation to
$|F_{N}(\nu )|^{2}$.  Throughout this work we apply the fractional RMS squared
normalisation,
\[
P(\nu)=\frac{2\,T}{\mu^{2}\,N^{2}}\:|F_{N}(\nu )|^{2},
\]
which is commonly used in measuring XRB power spectra and has the desirable
property that integrating the power spectrum over a given frequency range,
$\nu_{1}$ to $\nu_{2}$ yields the
contribution to the fractional RMS squared variability (i.e.
$\sigma^{2}/\mu^{2}$) of the lightcurve due to variations on
time-scales of $\nu_{2}^{-1}$ to $\nu_{1}^{-1}$ (e.g. van
der Klis 1997).  Thus the total
fractional RMS variability of the lightcurve is given by the square root
of the integral of the power spectrum across
all measured frequencies, $\nu_{\rm min}$ to $\nu_{\rm Nyq}$.  Under
this normalisation, the constant level of power contributed to all
frequencies by the Poisson noise in the lightcurve is equal to
$2(\mu+B)/\mu^{2}$, where $B$ is the mean background count rate.  Using
this normalisation allows us to compare power spectra measured by
different instruments and power spectra of different sources,
and take account of the linear RMS-flux relation recently
discovered in AGN and XRBs (Uttley \& M$^{\rm c}$Hardy 2001, and see
Section~\ref{station}). \\
For each source in our sample we have lightcurves for three observing 
schemes, which we use to measure power spectra over three different
frequency ranges to produce the broadband power spectrum:
\begin{enumerate}
\item[1.] A long-term monitoring lightcurve incorporating {\it all} monitoring
data, to measure the low-frequency power spectrum
($\sim10^{-8}$~Hz--$10^{-6}$~Hz).
\item[2.] An intensive monitoring lightcurve, to measure a medium-frequency
power spectrum ($\sim10^{-6}$~Hz--$10^{-5}$~Hz).
\item[3.] A long-look lightcurve (two such lightcurves for NGC~3516) to
measure the high-frequency power spectrum $\sim10^{-5}$~Hz--$10^{-4}$~Hz.
\end{enumerate}
Additionally, for the most variable sources MCG-6-30-15 and NGC~5506,
which show significant variability on time-scales less than 1~ks,
we measure a very-high-frequency (VHF) power spectrum
($\sim4\times 10^{-4}$~Hz--$10^{-2}$~Hz) using continuous
$\sim2.5$~ks segments of the PCA lightcurves (i.e. between
Earth-occultations of the source), binned to 16~s resolution.  We do not
include VHF power spectra for NGC~5548 and NGC~3516, since they show no
significant source power, other than the small amount expected at the lowest
frequencies due to red-noise leakage of variations
which are sampled by the high-frequency power spectrum (see
Section~\ref{alias}). \\
In order to minimise any distortion, the power spectra are made
from lightcurves binned up to the
maximum sampling interval of the observing scheme under consideration
(i.e. the lightcurve resolution is two weeks or 1209.6~ks for the long-term monitoring
lightcurves, 86.4~ks for the intensive monitoring lightcurves, except
for NGC~3516 where we bin the long-term and intensive monitoring
lightcurves to 4.3 days and 12.8 hours respectively).  Long-look
lightcurves are binned to 2048~s for the purposes of making the
high-frequency power
spectra, so that gaps due to Earth occultation are minimised to be no
more than one bin wide. 
Empty lightcurve bins in the binned-up monitoring and long-look lightcurves
 are filled by linearly interpolating
between adjacent filled bins.  No rebinning or interpolation was applied
to the 16~s lightcurves used to determine the VHF power spectra of
NGC~5506 and MCG-6-30-15, since only continuous sections of the
lightcurves were used to estimate the power spectrum. \\
The total lightcurve durations, bin widths, mean fluxes, fractional RMS
variability (after subtracting the Poisson noise contribution to variance) 
and power-spectral Poisson noise levels for each lightcurve are given in 
Table~\ref{tab:lctab}.  Note
that mean flux and fractional RMS are calculated based on the quoted bin
widths, i.e. the contributions to mean flux and
fractional RMS from each bin are equally
weighted, so that bins containing many data points (e.g. 2-week wide
bins which contain daily or twice-daily observations) do not contribute
more to the mean flux or variance than bins which contain a single data
point. \\
\begin{table*}
\caption{Parameters of lightcurves for use in {\sc psresp}} \label{tab:lctab}
\begin{tabular}{lccccc}
 & \multicolumn{5}{c}{Long-term} \\
& $T$ & $\Delta T_{\rm samp}$ & $\mu$ & $\sigma_{\rm frac}$ & $P_{\rm noise}$ \\
MCG-6-30-15 & $8.52\times10^{7}$ & $1.2096\times10^{6}$ & 14.0 & 26.5\% & 0.26 \\
NGC~5506 & $8.52\times10^{7}$ & $1.2096\times10^{6}$ & 26.8 & 22.6\% & 0.11 \\
NGC~5548 & $8.26\times10^{7}$ & $1.2096\times10^{6}$ & 13.6 & 30\% & 0.29 \\
NGC~3516 & $5.60\times10^{7}$ & $3.6864\times10^{5}$ & 13.3 & 29.6\% & 0.28 \\
& \multicolumn{5}{c}{Intensive} \\
& $T$ & $\Delta T$ & $\mu$ & $\sigma_{\rm frac}$ & $P_{\rm noise}$ \\
MCG-6-30-15 & $3.22\times10^{6}$ & $8.64\times10^{4}$ & 14.7 & 21.7\% & 0.24 \\
NGC~5506 & $3.61\times10^{6}$ & $8.64\times10^{4}$ & 22.4 & 15.1\% & 0.14 \\
NGC~5548 & $3.58\times10^{6}$ & $8.64\times10^{4}$ & 14.8 & 20\% & 0.26 \\
NGC~3516 & $1.18\times10^{7}$ & $4.608\times10^{4}$ & 12.7 & 28.7\% & 0.30 \\
& \multicolumn{5}{c}{Long-look} \\
& $T$ & $\Delta T$ & $\mu$ & $\sigma_{\rm frac}$ & $P_{\rm noise}$ \\
MCG-6-30-15 & $7.23\times10^{5}$ & 2048 & 12.2 & 20.8\% & 0.32  \\
NGC~5506$^{a}$ & $2.25\times10^{5}$ & 2048 & 6.9 & 12.1\% & 2.02 \\  
NGC~5548 & $1.97\times10^{5}$ & 2048 & 22.5 & 8.7\% & 0.14  \\  
NGC~3516$^{b}$ & $3.62\times10^{5}$ & 2048 & 11.5 & 7.2\% & 0.35 \\
NGC~3516$^{c}$ & $2.88\times10^{5}$ & 2048 & 15.2 & 9.8\% & 0.23 \\
\end{tabular}

\medskip
\raggedright{$T$ and $\Delta T_{\rm samp}$ are the lightcurve duration and sampling
interval (in seconds), $\mu$ and
$\sigma_{\rm frac}$ are the lightcurve mean flux (in count~s$^{-1}$) and
fractional RMS respectively and $P_{\rm noise}$ is the Poisson noise level
expected in the power spectrum due to counting statistics (in
fractional RMS-squared units, Hz$^{-1}$).  Notes:
$^{a}$~Details given in the table are for the {\it EXOSAT} lightcurve,
the {\it RXTE} lightcurve used to measure the power spectrum at the
highest frequencies has $\mu=28.1$, $P_{\rm noise}=0.10$.
$^{b}$~Lightcurve obtained 22--26~May 1997.
$^{c}$~Lightcurve obtained 13--16~April 1998.}
\end{table*}
\begin{figure*}
\begin{center}
{\epsfxsize 0.5\hsize
 \leavevmode
 \epsffile{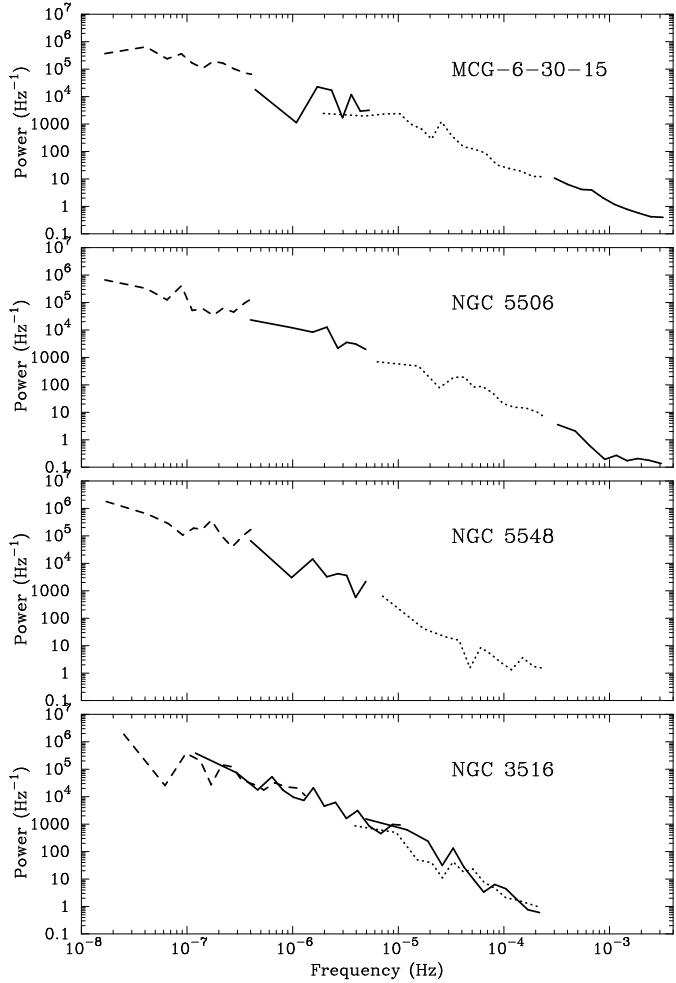}
}\caption{Raw broadband power spectra of MCG-6-30-15, NGC~5506, NGC~5548
and NGC~3516.  The dashed line shows the low-frequency part of the power
spectrum, made from the total monitoring lightcurves, while the dotted
line shows the high-frequency part made from long-look lightcurves. 
Solid lines mark the medium-frequency power spectrum (made from the
intensive monitoring lightcurves), VHF power spectra for MCG-6-30-15
and NGC~5506, and the power spectrum of the second long-look observation
for NGC~3516.  Poisson noise levels have not been subtracted from the
power spectra.} \label{rawpow}  
\end{center}
\end{figure*}
We measured each power spectrum using the method and normalisation
outlined above.  In order to reduce the scatter in power-spectral points,
which fluctuate
wildly for a stochastic process such as red or white-noise, 
we binned the logarithm of power at each frequency (see Papadakis \&
Lawrence 1993) in
logarithmically spaced frequency bins,
separated by a factor of 1.3 in frequency but with a minimum of
two measured powers per bin, so that the bin spacing is larger at the lowest
frequencies sampled by each power spectrum.  The VHF power
spectra for NGC~5506 and MCG-6-30-15 were calculated by measuring
separate power
spectra for each continuous lightcurve segment, averaging them and binning
in logarithmically spaced bins separated by a factor of 1.3 in
frequency.  The resulting
broadband power spectum for each object is shown in
Figure~\ref{rawpow}. \\
Inspection of the power spectra in Figure~\ref{rawpow} shows that they
do flatten at low frequencies.  However, we cannot immediately assume that this
flattening is real and representative of the shape of the true,
`underlying' power spectrum, for
the following reasons: 1) First of all, we cannot estimate reliable
errors for all but the VHF power spectra, especially for the points at
the low-frequency end of each power spectrum, due to the small number of points
which contribute to each frequency bin, 2) although rebinning and
interpolation result in evenly sampled lightcurves, they also introduce
distortions in the estimated power spectra which are difficult to
predict {\it a priori}.  Furthermore, even if these distortions are minimal,
the estimation of red-noise power spectra is affected by potentially serious
distortions due to aliasing and red-noise leak, which are dependent on the
original sampling pattern.  Finally, 3) we must consider the possibility
that the underlying power spectra are not stationary, but vary on time-scales
comparable to the length of our campaign, so that it is not valid to
combine power spectra taken at different times and over different intervals. 
We consider these problems in more detail in the remainder of this section.

\subsection{Error estimation}
\label{errors}
The smooth functions used to fit the power
spectra of noise processes such as red-noise, white-noise and the
composite broadband noise
represent the average power spectrum of the underlying noise
process, $P_{\rm proc}(\nu)$.  However the light curve which we measure is
a stochastic {\it
realisation} of that process and results in an observed power spectrum
$P_{\rm obs}(\nu)$
which fluctuates randomly about $P_{\rm proc}(\nu)$, following a $\chi^{2}$
distribution with 2 degrees of freedom and standard
deviation at any frequency $\nu$ equal to $P_{\rm proc}(\nu)$
(e.g. Timmer \& K\"{o}nig 1995). 
Therefore, in order to recover the underlying power spectrum of the
process $P_{\rm
proc}(\nu)$ directly from the data, we must determine the mean power
spectrum by averaging many observed power spectra, using the spread in the
observed power measured at each frequency to estimate the standard
error on the mean.  Because the $\chi^{2}$
distribution is exponential the power at a given frequency fluctuates
wildly, so the number of power spectra averaged
must be large ($>50$), in order that the standard error is reliable. 
This problem has been discussed extensively by Papadakis \&
Lawrence (1993), who show how more reliable estimates of smoothed power
(and the standard error) can be obtained by averaging fewer power
spectra ($\sim20$) if we instead average the logarithm of power rather
than the power.    \\
Unfortunately, we cannot use this method of error estimation to
constrain the shape of the power spectra we measure here (other than the
VHF power spectra), because there are not enough
data points to estimate reliable errors, especially at the lowest
frequencies measured in each power spectrum.
Therefore we must discard the requirement that the data are used to directly
estimate a reliable mean power-spectral shape and errors and instead use a 
Monte~Carlo technique, using simulated lightcurves to
estimate the power-spectral shape and uncertainty for
a range of specified models and test these against the data.  Using this
approach we can estimate reliable uncertainties even in the limit of
small-number statistics and hence use the full range of power-spectral
frequencies available to us.  Furthermore we can take account of the
distorting effects of lightcurve sampling, which we detail below.

\subsection{Power-spectral distortion due to lightcurve sampling}
\label{alias}
\begin{figure}
\begin{center}
{\epsfxsize 0.9\hsize
 \leavevmode
 \epsffile{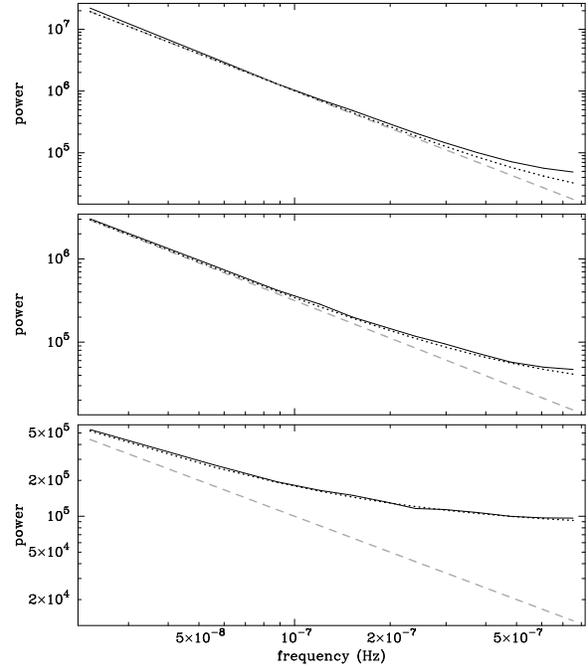}
}\caption{The effects of aliasing on power-spectral shape.  The
grey dashed line represents the true underlying power spectrum, while
the solid black line shows the average observed power spectrum after distortion
by the effects of discrete sampling at weekly intervals, 
obtained from averaging the power spectra of 1000 simulated lightcurves
of 2-year duration (see text).  The dotted line shows our
approximation to the average observed power spectrum, obtained by assuming that
the effect of aliasing is to add a constant to the underlying power
spectrum.  Underlying power spectral slopes
from top to bottom are $\alpha=$2.0, 1.5 and 1.0 respectively.}
\label{fig:alias}
\end{center}
\end{figure}

As we described in Section~\ref{powmeas}, the lightcurves
that we use to calculate the power spectra are rebinned to an even pattern
and any empty bins are filled with interpolated flux measurements.  The
use of the Monte~Carlo technique mentioned in the previous section can
take account of the distorting effects of rebinning and interpolation on
the power spectrum (see Section~\ref{lcsim}).  However, the estimation
of a red-noise power spectrum, even from an evenly sampled lightcurve,
is not free from distortions. \\
Consider an underlying continuous lightcurve $r(t)$ whose Fourier
transform is $R(\nu)$, on which we impose a sampling pattern $w(t)$ so that
$w(t)=1$ when we sample $r(t)$ and zero otherwise.  The resulting
observed lightcurve, $f(t)$ is given by:
\[
f(t)=r(t)w(t).
\]
Applying the Convolution theorem of Fourier transforms, the Fourier
transform of $f(t)$, $F(\nu)$ is then given by the convolution of
$R(\nu)$ and the Fourier transform of the sampling pattern $W(\nu)$
(known as the `window function'), i.e.
\[
F(\nu)=R(\nu)*W(\nu).
\]
Therefore the Fourier transform (and hence the power spectrum) of the
observed lightcurve is distorted from the true underlying power spectrum
by the sampling pattern imposed on the underlying lightcurve. 
Qualitatively we can distinguish two significant components to this
distortion in the case of red-noise power spectra, {\it red-noise leak}
and {\it aliasing}. \\
Significant power below the minimum frequency sampled by the
power-spectrum ($\nu_{\rm min}=T^{-1}$) causes long-time-scale trends in
the lightcurve which cannot be distinguished from smaller amplitude 
trends on the
time-scales which are sampled by the power spectrum.  The result
of this red-noise leak
is that additional power is transferred across the entire measured power
spectrum, with an
amplitude dependent on the amount of power at frequencies below $\nu_{\rm min}$
(and hence the amount of red-noise leak is model dependent and stochastic).  Fortunately, the effects
of red-noise leak can be accounted for using the Monte~Carlo technique
mentioned earlier, by ensuring that the simulated `underlying'
lightcurves are much longer than the observed lightcurves (see
Section~\ref{lcsim}).    \\
If a lightcurve is not continuously sampled (i.e. it is sampled for a
duration $\Delta T_{\rm bin}$ at sampling intervals $\Delta T_{\rm
samp}$, where $\Delta T_{\rm bin}\ll\Delta T_{\rm samp}$) then
variations on time-scales shorter than $\Delta T_{\rm samp}$ (i.e.
corresponding to power above the Nyquist frequency, $\nu_{\rm Nyq}$) 
cannot be distinguished
from (and therefore appear to contribute to) variations on longer time-scales. 
The result is that power is shifted or `aliased' to lower frequencies 
from frequencies above $\nu_{\rm Nyq}$.  Technically, the effect of
aliasing is to transfer the power at a frequency above the Nyquist
frequency, $\nu_{\rm Nyq}+\Delta \nu$ to a frequency below the Nyquist
frequency $\nu_{\rm Nyq}-\Delta \nu$, i.e. the power is reflected about
$\nu_{\rm Nyq}$ (e.g. see van der Klis 1989).  Hence, the amount and form of aliasing in the observed power spectrum
is dependent on the underlying power-spectral shape and amplitude. \\
However, since for all but the steepest broadband-noise type power
spectra, the power at $2\nu_{\rm Nyq}$ is not much less than the
power at $\nu_{\rm Nyq}$,
the result of aliasing can be approximated by adding a constant level of
power to the underlying power spectrum. 
For lightcurves with initial time resolution $\Delta T_{\rm bin}$ (prior to any
rebinning, e.g. 1~ks in the case of our monitoring lightcurves), we
expect that variations with frequencies higher than
$\sim(2\Delta T_{\rm bin})^{-1}$ will be smoothed out and will not
contribute significantly to aliasing.  For this reason, we expect
the total {\it integrated} power transferred to the
observed power spectrum by aliasing to be roughly equal to the integrated
power between the Nyquist frequency and $(2\Delta T_{\rm bin})^{-1}$. 
As a first approximation, we assume that this power will be distributed
evenly to all sampled frequencies, with the constant power, $P_{\rm C}$,
added to all frequencies because of aliasing given by:
\[
P_{\rm C}=\frac{1}{\nu_{\rm Nyq}-\nu_{\rm min}} \int_{\nu_{\rm
Nyq}}^{(2\Delta T_{\rm bin})^{-1}} P(\nu)\,d\nu.
\]
The effects of aliasing on power spectra of
different underlying slopes is shown
in Figure~\ref{fig:alias}.  The average aliased power spectra were
constructed from 1000 simulated lightcurves, each lightcurve
corresponding to 2 years of evenly spaced weekly 1~ks snapshots.  The
assumed underlying power
spectrum (shown by the dashed lines) was cut off above $10^{-8}$~Hz, to
reduce the red-noise leak
contribution to the power spectra.  We also plot our estimate of the 
aliased power spectrum
obtained using the constant-power approximation outlined above (dotted lines). 
Note that the flattening of the
power-spectral slope is more pronounced for flatter underlying slopes,
as expected.  At high
frequencies, the agreement between the aliased power spectrum and our
estimate is very good, except for the steepest power spectrum where the
amount of aliasing is small anyway.  \\
Fortunately, the distorting effects of aliasing can be properly taken account
of using the Monte~Carlo technique mentioned earlier.  The approximation
to the distorting effect of aliasing presented here will prove useful for
reducing
the resolution of the simulated lightcurves required by the technique,
as we will show in Section~\ref{lcsim}.

\subsection{Stationarity of the power spectrum}
\label{station}
A key assumption we must make, in order to combine lightcurves obtained at
different times to measure the broadband power spectrum, is that the
underlying power spectrum is stationary, i.e. its amplitude and shape 
do not change over time (note that power-spectral non-stationarity is not 
the same as
lightcurve non-stationarity, which is expected on short time-scales
for red-noise processes).  If the underlying power spectrum were non-stationary,
so that it happened to be steeper while the high-frequency power spectrum was
measured than when the medium or low frequency power spectra were
measured, then these changes in power-spectral slope would masquerade as a
flattening in the power spectrum.  \\
The power spectra of BHXRBs are known
to be non-stationary on a variety of time-scales, showing drastic
changes in shape between the well-known low and high
states (e.g. Cui et al. 1997).  Transitions between these states
occur on time-scales
of months to years and less drastic changes in power-spectral shape 
within the low and high states occur on time-scales of days or weeks
(e.g. changes in the high-frequency slope and the
position of the power-spectral breaks,
Belloni \& Hasinger 1990, Cui et al. 1997).  If the variability time-scales
for this kind of
non-stationarity scale linearly with black hole mass, we would 
expect to see similar changes in the power spectra of AGN on time-scales
of centuries or longer - much greater than the time-scale we can sample.
 This picture is supported by the fact that, to date, no hard evidence
for non-stationarity in power-spectral shape has been reported for AGN.
 Hence it is probably safe to assume that the shapes of our target's 
power spectra do not vary during the course of our monitoring campaign.\\
An alternative concern is that the amplitude of the power spectrum
varies over time, even though the shape does not.  In a separate paper,
we report the intriguing (and unexpected) result that the X-ray RMS
variability of Cyg~X-1 and the accreting millisecond pulsar
SAX~J1808.4-3658 scale linearly with the local mean flux (Uttley \&
M$^{\rm c}$Hardy 2001).  In other
words, the fractional RMS (i.e. RMS divided by mean flux) measured
for any fixed-length segment of the lightcurve is constant
(subject to the stochastic variability inherent in the lightcurves),
irrespective
of the segment's mean flux.  We also confirm that the same property applies
to the lightcurves of AGN and hence appears
to be an important characteristic of the broadband noise variability in compact
accreting systems.  It is obvious then that we must always normalise our
lightcurves by their mean before calculating their power spectra or,
as we do here, divide their power spectra by the square of the mean,
otherwise the shape of our broadband power spectrum will be dependent on the
mean object flux at the time we measured each lightcurve.

\section{{\sc psresp}: a reliable method of estimating power-spectral shape}
\label{psresp}
In order to take proper account of the errors on the power spectrum and the
distorting effects due to rebinning, red-noise leak and
aliasing, we must use
a Monte~Carlo technique to estimate the underlying power-spectral shape
(and its uncertainty).  The basic concept is to simulate a large number
of continuous lightcurves with a known underlying power-spectral shape, apply the
sampling pattern of the observed lightcurve, rebin and interpolate as
necessary, obtain the resulting
distorted power spectra and average them to determine the mean shape of the distorted
model power spectrum.  The spread of the simulated power spectra about
the mean can be used to estimate the errors on the observed power
spectrum. Applying the sampling pattern to a simulated (i.e. model)
lightcurve before measuring its power spectrum is
equivalent to convolving the window function of the sampling pattern with
the Fourier transform of the model lightcurve to obtain the
Fourier transform of the observed lightcurve.  The effect is then similar to
that used in X-ray spectral measurement, when a model spectrum is
postulated and then convolved with the `response function' of the
detector to yield the `observed' model spectrum which can be tested
against the data.  Hence, this Monte-Carlo technique for estimating the
true power-spectral shape is known as the `response method'. \\
The response method was applied by Done et al. (1992), to measuring
the power spectrum of unevenly sampled {\it Ginga} lightcurves and has
subsequently been applied to {\it ROSAT} data (Green, M$^{\rm c}$Hardy
\& Done 1999).  In both cases the method was applied only to a single
lightcurve, of relatively short (a few days) duration.  Here, however,
we wish to fit the same power-spectral model to several power spectra in
order to estimate the shape of the broadband power spectrum.  We should
not fit power spectra measured over different frequency ranges
separately, because the distorting effects of aliasing and red-noise
leak, which we must take account of, are dependent on the shape of the
underlying power spectral shape outside the frequency range measured by a
single power spectrum. 
Furthermore, the response method of Done et al. (1992) estimates a goodness of
fit based on the standard $\chi^{2}$
statistic, which is not a reliable estimator for the distorted power
spectra considered here and hence cannot be used to formally estimate
fit probabilities.  Therefore, for our purpose of reliably constraining
the shape of the broadband power spectrum of AGN, we have developed a
more sophisticated technique based on the response method, which we call
{\sc psresp} (for {\sc p}ower {\sc s}pectral {\sc resp}onse).  In the
remainder of this section, we describe {\sc psresp} in detail.

\subsection{Lightcurve simulation}
\label{lcsim}
We simulate lightcurves using the method of Timmer \& K\"{o}nig (1995),
which is superior to the commonly used method of summing sine waves with
randomised phases in that the power-spectral amplitude at each frequency
is randomly drawn from a $\chi^{2}$ distribution, as should be 
the case for noise processes, and not fixed at the amplitude of the underlying
power spectrum.  We create the fake `observed' lightcurves as follows. \\
First, we specify a power-spectral model which we wish to test against
the data (which is some continuous function such as a power-law, with or
without breaks).  The normalisation of the model power spectrum is a 
multiplicative factor which is carried
through any convolution with the window function (i.e. only the
power-spectral shape is distorted by sampling).  Hence we can choose an
arbitrary model normalisation and simply renormalise the 
resulting distorted model power spectrum to obtain the best possible fit
to the data. \\
For each observed lightcurve used to measure the broadband power
spectrum, we simulate $N$ continuous lightcurves using the given 
power-spectral
model (where $N$ is large, between 100 and 1000).  Ideally, the time
resolution of the simulated lightcurves, $\Delta T_{\rm sim}$ should be
the same as the initial resolution of the observed lightcurve $\Delta T_{\rm
bin}$, which for our
monitoring data is the typical exposure time of the snapshot observations. 
This is because any variations on time-scales down to this resolution
will contribute to aliasing. Although we have
derived an analytical approximation to the effect of
aliasing on the power spectrum, this only tells us the {\it average} effect of
aliasing for lightcurves which are evenly sampled.  Due to the
stochastic nature of the lightcurves, the actual power above the Nyquist
frequency is variable and hence aliasing contributes a variable amount
to the power spectrum which adds to the uncertainty in its shape. However,
for the monitoring lightcurves, which
consist of only 1~ksec snapshots but may be months or years long,
the requirement that the simulated lightcurves also have 1~ksec 
resolution increases the computation time prohibitively.  \\
In practice, we can limit the resolution of the simulated lightcurves to be
$\Delta T_{\rm sim}\leq0.1\Delta T_{\rm samp}$. 
This is because, for lightcurves whose power spectra are steep at
high frequencies, like those we measure here, the uncertainty in the
amount of aliasing is dominated by the uncertainty in the large amount
of power at frequencies not much greater than the Nyquist frequency. 
We can then estimate the much smaller
contribution to aliasing due to power at frequencies greater than
sampled by the simulated lightcurves using our analytical approximation, i.e.
calculating the integrated power of the model power spectrum
between $(2 \Delta T_{\rm sim})^{-1}$ and $(2 \Delta T_{\rm bin})^{-1}$
(see Section~\ref{alias}) and incorporating the resulting constant values
into the final simulated power spectra. \\
In order to take
account of red-noise leak, the simulated lightcurves must be
significantly longer (e.g. by a factor 10 or more) than the observed
lightcurve, so that there is power at frequencies lower than the minimum
frequency sampled by the observed lightcurve.  We can minimise the cost
of simulating excessively long lightcurves by making a single, very long
lightcurve of length $NT$, which is subsequently split into the $N$ desired
segments.  Note that although the longest
time-scale trends in the total simulated lightcurve contribute the
same amount to the red-noise leak in the power spectra of all $N$ simulated
lightcurve segments, the bulk of red-noise leak (and the power-spectral
uncertainty it introduces) is due to variations on shorter time-scales
close to the observed lightcurve duration and hence will be
statistically independent between segments. \\
Once a continuous lightcurve is simulated, it is resampled using the
same sampling pattern as the observed lightcurve.  The resampled
lightcurve is then rebinned and empty bins interpolated in the same
manner as for the observed lightcurve.  The power spectrum of
the resulting lightcurve is then measured.

\subsection{Determining the goodness of fit of the model}
\label{goodfit}
The power spectrum of each simulated lightcurve is calculated and
binned in
the same way as the power spectrum of the original observed lightcurve. 
The model average power spectrum (corresponding to the given model and
sampling pattern of the observed lightcurve) is then given by the mean 
of the $N$ simulated power spectra, $\overline{P_{\rm sim}}(\nu)$. 
For each frequency $\nu$ sampled by the
binned power spectrum, we determine the RMS spread of simulated model
powers about the mean and define this spread to be the `RMS error' on the power
at that frequency, $\Delta \overline{P_{\rm sim}}(\nu)$.  We next define
a statistic, which we call $\chi^{2}_{\rm dist}$, which is calculated
from the model average (and RMS error) and observed power spectra of
each lightcurve:
\[
\chi^{2}_{\rm dist}=\sum_{\nu = \nu_{\rm min}}^{\nu_{\rm max}}
\frac{\left(\overline{P_{\rm sim}}(\nu)-P_{{\rm obs}}(\nu)\right)^{2}}{\Delta
\overline{P_{\rm sim}}(\nu)^{2}},
\]     
where $\nu_{\rm min}$ and $\nu_{\rm max}$ are respectively the
minimum and maximum frequencies measured by $P_{{\rm obs}}(\nu)$. 
We measure $\chi^{2}_{\rm dist}$ for each input power spectrum (i.e. low,
medium and high-frequency), and sum to
yield a total $\chi^{2}_{\rm dist}$ for the
model with respect to the data.  Note that although the approach of
assigning error bars to the model rather than the data is unusual, it is
technically valid and strictly speaking the more correct thing to do,
since the $\chi^{2}$ statistic is defined on the basis of
the variance of the model population, which error bars conventionally
determined from the data are supposed to approximate (e.g. see
discussion in Done et al. 1992, Section~4.2). \\
Next, we find the best-fitting normalisation of the power-spectral model
by renormalising the $\overline{P_{\rm sim}}(\nu)$ for each input power
spectrum by the same factor $k$, varying $k$ until the total
$\chi^{2}_{\rm dist}$ is minimised. \\
Having determined the minimum $\chi^{2}_{\rm dist}$ for the model
compared to the data, we next estimate what goodness of fit this
value of $\chi^{2}_{\rm dist}$ corresponds to.  The $\chi^{2}_{\rm
dist}$ is {\it not} the same as the standard $\chi^{2}$ distribution,
because the $P_{\rm obs}(\nu)$s are not normal variables (since the
number of power spectrum estimates averaged in each frequency bin is small). 
Therefore we must estimate a reliable
goodness of fit using the distribution of $\chi^{2}_{\rm dist}$ and not
the well-known $\chi^{2}$ distribution.  For each input low, medium and
high-frequency power spectra, we
have already created $N$ corresponding simulated power spectra which can be used to 
determine the
distribution of $\chi^{2}_{\rm dist}$ for that particular model and
lightcurve sampling pattern.  In order to determine the distribution of
$\chi^{2}_{\rm dist}$, we should calculate the values of $\chi^{2}_{\rm
dist}$ for all possible combinations of simulated low, medium and high-frequency
power spectra.  However, since the number of such combinations ($N^{3}$) may be
extremely large, we reduce the
number of combinations we sample to $M$ (where $M\ge1000$), randomly 
selected to ensure an accurate estimate of the distribution of
$\chi^{2}_{\rm dist}$. \\
We sort the $M$ simulated 
measurements of the total $\chi^{2}_{\rm dist}$ into ascending order.  {\it The
probability that the model can be rejected is then given by the
percentile of the simulated $\chi^{2}_{\rm dist}$ distribution
above which $\chi^{2}_{\rm dist}$ exceeds that measured for the
observed power spectra}.  Note that this method of using simulated data
sets to estimate goodness of fit in the absence of a well-understood
fit statistic is well-known and described in Press et al. (1992),
Section~15.6. 

\subsection{Incorporating VHF power spectra}
\label{vhfpow}
As described in Section~\ref{powmeas}, we use continuous segments of the
16~s binned long-look lightcurves to make VHF power spectra for our most
variable targets, by averaging the power spectra of all the segments and
determining the standard error from the spread in power at each
frequency.  The standard errors
estimated from the data are reliable, since we average $>20$ power
spectra and bin the logarithm of power (according to the method of
Papadakis \& Lawrence 1993 and see Section~\ref{errors}).   We can therefore
use the
measured VHF power spectra and their errors as they are, without having to
estimate errors using simulations of high-resolution lightcurves which
would be very computationally intensive.
However, if we simply compare the VHF
power spectrum with the underlying undistorted model shape, we ignore
the effects of red-noise leak which could be significant in distorting
the shape of the VHF power spectrum, especially if the underlying power
spectrum does not flatten significantly until far below the minimum
frequency sampled by the VHF power spectrum.  Therefore, we
need to take account of the effects of red-noise leak on the model
power-spectral shape at high frequencies.  Note that, because the
segments are continuous and binned into high-resolution time bins of
width 16~s, the effects of aliasing are not important in this case. \\
To determine the effects of
red-noise leak on the VHF power spectrum, we simulate 1000 lightcurves, each
made to be at least 64 times longer than
$\nu_{\rm min}^{-1}$ (where $\nu_{\rm min}$ is
the minimum frequency sampled by the VHF power
spectrum), with resolution $\Delta T_{\rm sim}$
smaller than $\frac{1}{2}\nu_{\rm
max}^{-1}$ where $\nu_{\rm max}$ is the maximum frequency which
contains significant power above the noise level (typically around
$3\times10^{-3}$~Hz) and is chosen so that the ratio of $\nu_{\rm max}$
to
$\nu_{\rm min}$ is a power of 2.  Power spectra of the
lightcurves sampled to have duration $\nu_{\rm min}^{-1}$ may then be
made using the Fast Fourier Transform, which allows a VHF model average
power spectrum for 1000 simulated lightcurves to be
determined very rapidly.  The VHF model average power spectrum
is then used in place of the underlying model power spectrum,
while the errors determined from the
observed VHF power are used as errors on the model. \\
The contribution of the VHF power spectrum to the total $\chi^{2}_{\rm
dist}$ is determined by comparing the observed power spectrum with the
simulated model average power spectrum, using the standard errors
estimated from the data.  The contribution of the VHF power to the
goodness of fit of the model
is obtained as follows:  for each of the $M$ random combinations of
simulated power spectra used to estimate the goodness of fit, a VHF
power spectrum is simulated by randomly selecting the power at each
measured frequency, from a Gaussian distribution of mean equal to the
model average power and standard deviation equal to the standard error
at that frequency.  The $\chi^{2}_{\rm dist}$ of the simulated VHF power
spectrum is determined and included in the total $\chi^{2}_{\rm dist}$
measured for that selection of simulated power spectra.  The goodness of
fit of the model is then estimated as described in the preceding section.

\subsection{Summary of the {\sc psresp} method}
We summarise the {\sc psresp} method as follows:
\begin{enumerate}
\item[1.] Measure the observed power spectrum of each (rebinned and 
mean-subtracted)
observed lightcurve and bin up the power spectrum as desired
(see Section~\ref{powmeas}). 
Measure the VHF power spectrum if required, determining its errors
directly from the data (see Section~\ref{vhfpow}).
\item[2.] Specify the underlying power spectral model to be tested against
the data.  For the given set of parameters, simulate $N$ lightcurves
which are realisations of the model and apply the sampling pattern of
the observed lightcurve to the simulated lightcurves (see
Section~\ref{lcsim}).
\item[3.] Calculate the
power spectrum of each re-sampled simulated lightcurve using the same
method used to measure the observed power spectrum.  Determine the model
average power spectrum from the $N$ simulated power spectra, and assign
error bars equal to the RMS spread in simulated power at each frequency
(see Section~\ref{goodfit}). \\
 The two steps above should be repeated for each lightcurve (i.e.
long-term, intensive and long-look), to make simulated model average power
spectra corresponding to each observed power spectrum.  The model
average VHF power spectrum should also be determined at this point (if
required), according to the method outlined in Section~\ref{vhfpow}.
\item[4.] Estimate the statistic $\chi^{2}_{\rm dist}$ (summed over all
input power spectra) for the observed versus
the model average power spectrum, and vary the normalisation of the model
to minimise $\chi^{2}_{\rm dist}$ and obtain the best-fitting
normalisation (include the VHF power spectrum in this determination if
required, using standard errors determined from the data, see
Section~\ref{vhfpow}).  
\item[5.] Determine the $\chi^{2}_{\rm dist}$
for $M$ randomly selected combinations of the simulated
power spectra.  If a VHF power spectrum is included, measure
its contribution to each simulated total $\chi^{2}_{\rm dist}$ from a 
random realisation of
the model average VHF power spectrum, according to the method described
in Section~\ref{vhfpow}.  Sort the simulated distribution of
$\chi^{2}_{\rm dist}$ into increasing numerical order - the model is
rejected at a confidence equal to the percentile of the simulated
$\chi^{2}_{\rm dist}$ distribution above which $\chi^{2}_{\rm dist}$
exceeds that measured for the observed power spectra (see
Section~\ref{goodfit}).
\end{enumerate}
Using the method described above, any given model can be tested against
the data.  By stepping through a range of model parameters and
repeating steps 2--5, large regions of the model parameter space may be
searched and confidence regions may be determined. 

\section{Results}
\label{results}
 We will now apply the {\sc psresp} method described in Section~\ref{psresp} 
to the lightcurves of our sample, in order to
determine if the broadband power spectra of Seyfert galaxies really
flatten towards low frequencies and to try to constrain models for any
flattening which we see.

\subsection{Do the broadband power spectra really flatten?}
\begin{figure*}
\begin{center}
{\epsfxsize 0.5\hsize
 \leavevmode
 \epsffile{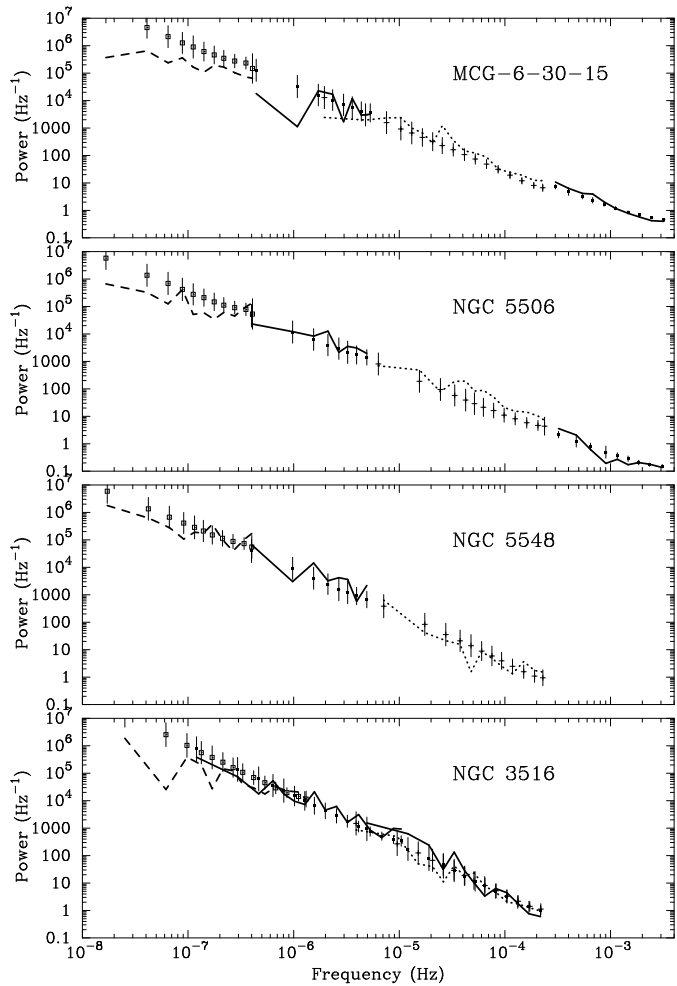}
}\caption{Comparison of best-fitting model average power spectra with
the observed power spectra for the single power-law model described in
the text.  Open squares mark the low-frequency model average, simple
crosses mark the high-frequency model average while filled squares mark
the medium-frequency model average, the VHF model average for
MCG-6-30-15 and NGC~5506, and the high-frequency model average for the
second long-look observation of NGC~3516.  Note that the error bars
represent the RMS error in the simulated power spectra used to calculate
$\chi_{\rm dist}^{2}$ as described in Section~\ref{goodfit}.} \label{nobkpow}  
\end{center}
\end{figure*}

\begin{figure}
\begin{center}
{\epsfxsize 0.9\hsize
 \leavevmode
 \epsffile{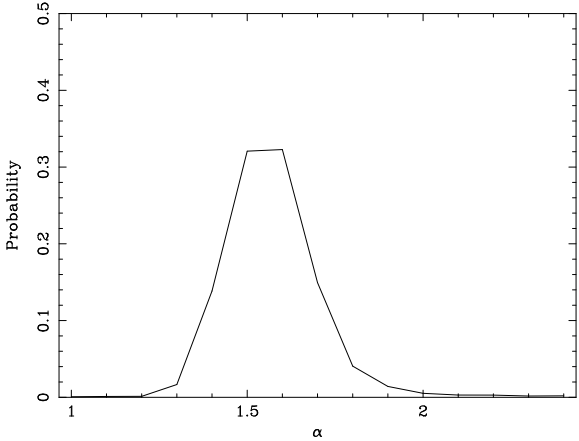}
}\caption{Probability that a single power-law of slope $\alpha$ is
acceptable to describe the broadband power spectrum of
NGC~5548.} \label{n5548nobkprob}
\end{center}
\end{figure}

To determine if the power spectra flatten, we will test a
simple power-law model for the underlying power spectrum, $P_{\rm
mod}(\nu)$ of the form:
\[
P_{\rm mod}(\nu)=A\,\left(\frac{\nu}{\nu_{0}}\right)^{-\alpha} + C_{\rm noise},
\]
where $A$ is the amplitude of the model power
spectrum at a frequency $\nu_{0}$, $\alpha$ is the power-spectral
slope and $C_{\rm noise}$ is a constant value which is fixed at the
Poisson noise level for the lightcurve.  Note that the Poisson noise
level is included in the model rather than subtracted from the power
spectra before model fitting, because the power spectra are binned
logarithmically (so constants in linear space may not simply be
subtracted).  This is particularly important for the VHF
power spectra of NGC~5506 and MCG-6-30-15, whose standard errors are
determined in logarithmic space, and also for high-frequency power
spectra in general, which are close to the Poisson noise level, since
fluctuations in the power spectrum lead to some measured powers lying
below the Poisson noise level (so subtraction of this level would lead
to negative measured powers).   \\
The model is tested against the measured power spectra
by stepping through values of $\alpha$ from 1.0 to 2.4
in increments of 0.1 (i.e. test the model with $\alpha=1.0$, 1.1, 1.2
etc.).  These values of $\alpha$ cover the range of reasonable values
which could possibly be fitted to the data.  Probabilities that the
measured power spectra could be a realisation of the model are
calculated by {\sc psresp}, as described in Section~\ref{goodfit}
using $N=1000$ simulated lightcurves to determine the distorted model
average power spectrum.  The distribution of $\chi^{2}_{\rm dist}$ of
the realisations of the model is determined for each value of $\alpha$
by measuring $\chi^{2}_{\rm dist}$ for $M=10^{4}$ randomly selected sets of
simulated power spectra.  The simulated lightcurves have time
resolutions $\Delta T_{\rm sim}$ given in Table~\ref{lcres}.  Additional 
distortion in the power spectrum due to model power at frequencies
greater than $(2\Delta T_{\rm sim})^{-1}$ is calculated directly from the
model, as described in Section~\ref{lcsim}.  Distorted VHF model power
spectra, which take account of red-noise leak in the VHF power spectra
included in the broadband power spectra of MCG-6-30-15 and
NGC~5506, are determined using the method described in
Section~\ref{vhfpow}. \\
\begin{table*}
\caption{Time resolution of simulated lightcurves used in {\sc psresp}.} \label{lcres}
\begin{tabular}{lccc}
& Long-term & Intensive & Long-look \\
MCG-6-30-15 & 86400~s & 8640~s & 512~s \\
NGC~5506 & 86400~s & 8640~s & 512~s \\
NGC~5548 & 86400~s & 8640~s & 512~s \\
NGC~3516 & 46080~s & 4608~s & 512~s \\
\end{tabular}
\end{table*}
The best-fitting values of $\alpha$, and corresponding confidences that the
single power-law model is rejected by the data,
are given in Table~\ref{nobkprob}.  The first and second
of each of these values shown for NGC~5506 correspond to fits without
or including the {\it EXOSAT} data respectively.  The simple power-law
model, without any flattening is rejected at better than
99\% confidence for MCG-6-30-15, better than 90\% confidence
(or close to 99\% confidence including the {\it EXOSAT} data) for
NGC~5506 and better than 95\% confidence for NGC~3516.  Only
for NGC~5548 is the model not rejected at a
significant confidence.  The best-fitting models are compared with the
measured power spectra in Figure~\ref{nobkpow}.  \\
It is apparent from
these plots that the simple power-law model does not fit the observed
power spectra of MCG-6-30-15, NGC~5506 and NGC~3516, even after allowing
for the distorting effects of sampling, because the intrinsic
power spectrum of each of these objects does indeed flatten towards low
frequencies.  There is no significant evidence for flattening at low
frequencies in the power spectrum of NGC~5548.  Figure~\ref{n5548nobkprob} shows the
fit probability measured at each input value of $\alpha$ for NGC~5548, which
demonstrates how {\sc psresp} is capable of finding well-defined 
probability maxima in the same way that $\chi^{2}$ fitting can, using
more conventional data sets. \\
\begin{table}
\caption{Results from fitting broadband power spectra of four Seyfert
galaxies with a simple unbroken power-law model.} \label{nobkprob}
\begin{tabular}{lcc}
& Best-fitting $\alpha$ & Rejection confidence \\
MCG-6-30-15 & 1.5 & 99.8\% \\
NGC~5506 & 1.4/1.5 & 90.6\%/98.6\% \\
NGC~5548 & 1.6 & 67\% \\
NGC~3516 & 1.8 & 96.6\% \\
\end{tabular}

\medskip
\raggedright{Quoted values for NGC~5506 correspond to fits
excluding/including the {\it EXOSAT} data.} 
\end{table}
The simple power law model with no frequency breaks can be rejected
at better than 95\% confidence for all objects except NGC~5548.  The next
step is to try to fit the observed power spectra with more complex models which
flatten at low frequencies, in particular, can we distinguish between
models where the power spectrum flattens to $\alpha=0$ or $\alpha=1$, and
can we constrain any characteristic frequencies for the flattening?

\subsection{Fitting simple models for the power-spectral flattening}
Although we can say with confidence that the power spectra of three of
the objects in our sample flatten towards low frequencies, it is not
clear what form this flattening takes.  In this work, we will restrict
ourselves to considering two simple models, a `knee' model based on the
low-frequency flattening seen in the power spectra of BHXRBs in the low
state (i.e. the low-frequency break in Cyg~X-1) and a `high-frequency
break' model which assumes that the flattening is due to a frequency
break in the power spectrum analogous to the high-frequency break seen
in the power-spectrum of Cyg~X-1.  Under the knee
model, the power-spectrum turns over to a slope $\alpha=0$ below some
`knee frequency', whereas under the high-frequency break model, the
power spectrum breaks sharply to $\alpha=1$ below some `break
frequency'. 
More complex models, consisting of
multiple frequency-breaks and a variety of power-spectral slopes, or
a number of broad Lorentzians (e.g. Nowak 2000) might also successfully 
explain the
data.  However, computational constraints limit the testing of a
large variety of
models for the flattening we see and moreover, as we shall discover, the data
do not yet warrant these kinds of models as the observed power spectra 
can be fitted adequately by the simple models we
test here.  We will fit these two simple models for the flattening to
the power spectra of all the Seyfert galaxies in our sample,
including NGC~5548 so as to set upper limits on any knee or break
frequencies. 

\subsubsection{The knee model}
\begin{figure}
\begin{center}
{\epsfxsize 0.9\hsize
 \leavevmode
 \epsffile{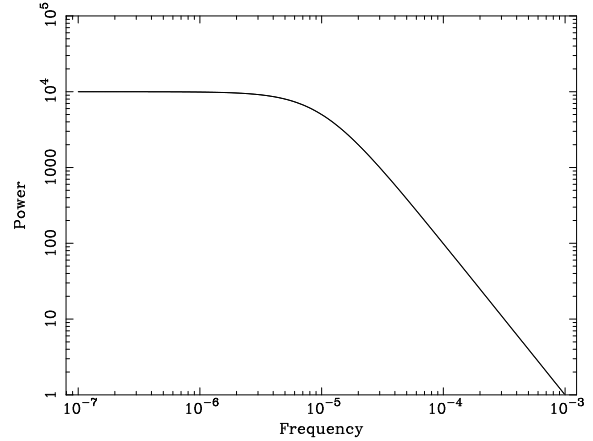}
}\caption{The form of the knee model power spectrum.} \label{kneemod}
\end{center}
\end{figure}

We first test the knee model for the power spectrum, which has the form
\[
P(\nu)=\frac{A}{\left(1+\left(\frac{\nu}{\nu_{\rm knee}}\right)^2\right)
^{\alpha /2}},
\]
where $A$ is the constant amplitude of the power-spectrum at zero slope,
$\nu_{\rm knee}$ is the `knee frequency' and $\alpha$ is now defined
as the power-spectral slope above the knee frequency.  The shape which
this function describes can be seen in Figure~\ref{kneemod}. 
We now test this model against the measured broadband power spectra,
to see if it can explain the flattening we see.  Using the equation
given above for the underlying power spectral shape (also including the
constant Poisson noise level), we can fit the model in the same way as
fitting a simple power law in the previous section.  The free parameters to
be stepped through are $\alpha$, which is again incremented in steps of
0.1 between 1.0 and 2.4, and $\nu_{\rm knee}$ which is stepped through
by multiplicative factors of 2, from $10^{-8}$~Hz to $10^{-3}$~Hz,
since a very broad range in frequency must be covered.   Approximately 200
pairs of $\alpha$ and $\nu_{\rm knee}$ must be tested (as opposed to
only 15 parameters when fitting the simple power law in the previous
section), so to save on computing time, the number of lightcurve 
simulations used to estimate each model average power spectrum
for each pair of parameters is reduced from $N=1000$ to $N=100$.  The
$\chi^{2}_{\rm dist}$ distribution is obtained by
determining $\chi^{2}_{\rm dist}$ for $M=1000$ sets of simulated power
spectra. \\
The best-fitting parameters and probabilities are shown in
Table~\ref{kneeprob}.
\begin{table*}
\caption{Results from fitting broadband power spectra of four Seyfert
galaxies with a knee model} \label{kneeprob}
\begin{tabular}{lccc} \\
& Rejection confidence & $\alpha$ & $\nu_{\rm knee}/10^{-6}$~Hz \\
MCG-6-30-15 & 81\% & $1.8\pm0.1$ & 5.12 (2.56--10.24) \\
NGC~5506$^{a}$ & 42\% & $1.7\pm^{0.7}_{0.3}$ & 0.64 (0.0--10.24) \\
NGC~5506$^{b}$ & 32\% & $1.9\pm^{0.5}_{0.4}$ & 2.56 (0.16--10.24) \\
NGC~5548 & 53\% & $1.6\pm^{0.8}_{0.2}$ & 0.02 (0.0--1.28) \\
NGC~3516 & 83\% & $2.1\pm0.3$ & 0.64 (0.32--1.28) \\
\end{tabular}

\medskip
\raggedright{The table shows the confidence that the knee model can be rejected, the
best fitting slope above the knee $\alpha$ and the best-fitting knee
frequency $\nu_{\rm knee}$ and 90\% lower and upper confidence limits to the
knee frequency (in brackets).
Errors quoted for $\alpha$ correspond to the
values of $\alpha$ below which the fit probability is reduced to less
than 10\% (i.e. they represent 90\% confidence limits). 
Note that these confidence limits are not interpolated
between sampled points in the parameter space, unlike the confidence
contours plotted in Fig.~\ref{kneepow}. 
The superscripts $^{a}$ and $^{b}$ mark the NGC~5506 results excluding
and including the {\it EXOSAT} data respectively.}
\end{table*}
\begin{figure*}
\begin{center}
{\epsfxsize 0.7\hsize
 \leavevmode
 \epsffile{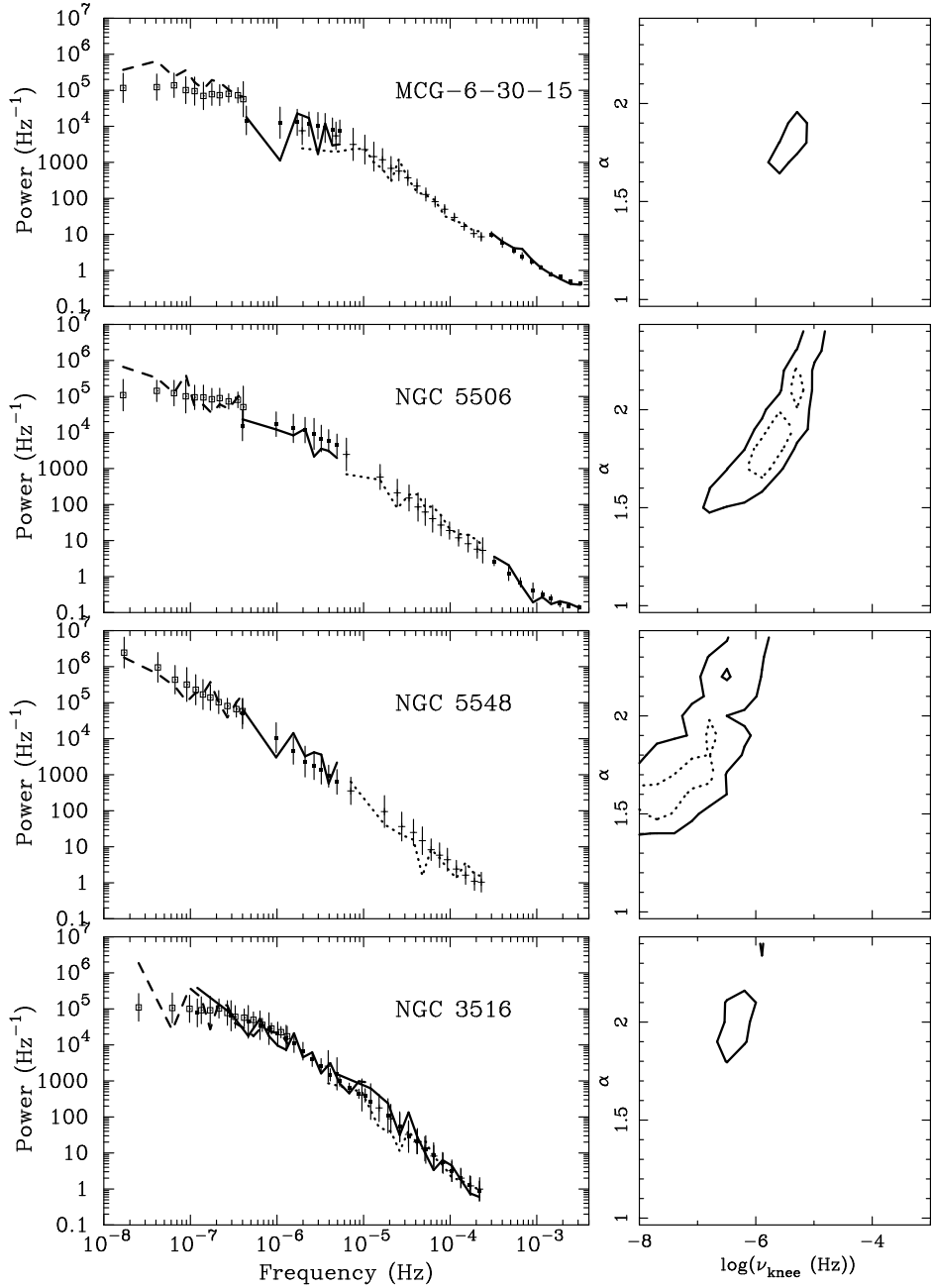}
}\caption{Comparison of best-fitting model average power spectra (points
with error bars) with
observed power spectra (lines) for the knee model described in the text (left),
and corresponding confidence contours for the parameter space searched
(right).  The dashed and solid confidence contours represent the 68\% and
90\% confidence limits respectively.} \label{kneepow}
\end{center}
\end{figure*}
Contour plots for each of the
knee model fits (not including the fit of the NGC~5506 broadband power
spectrum which excludes the {\it EXOSAT} data), together with the
best-fitting model average power spectra, are shown in
Figure~\ref{kneepow}.  The contour plots show that the acceptable
regions are broad and well-defined, rather than consisting of very many
separate `islands', which implies that using only
100 simulated lightcurves per measured power spectrum
is sufficient to determine reliable maxima in the probability space.  \\
As Table~\ref{kneeprob} shows, the knee model fits the power spectra of
all the objects in our sample adequately.  In all cases, the model can
be accepted at a confidence level better than 10\%.  Note that the broadband
power spectrum of NGC~5548 is consistent with this model, although the
lower confidence limits on the knee frequency cannot be defined,
in agreement with the acceptable simple power law fit to these data.
The fact that the knee model provides a good fit to the broadband power
spectrum of NGC~5506 including the {\it EXOSAT} data, is consistent 
with the power spectrum of NGC~5506 being stationary over 
time-scales as long as a decade. 

\subsubsection{The high-frequency break model}
\label{hfbk}
\begin{figure}
\begin{center}
{\epsfxsize 0.9\hsize
 \leavevmode
 \epsffile{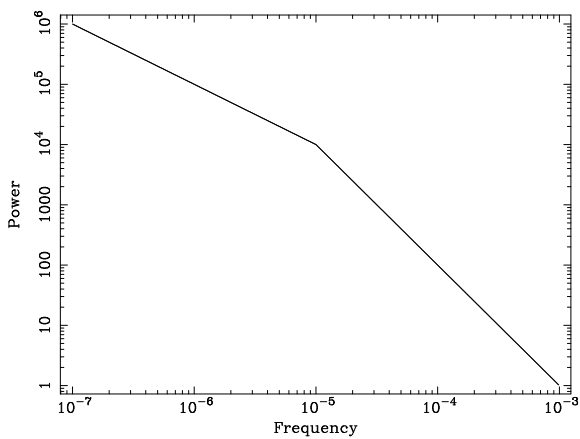}
}\caption{The form of the high-frequency break model power
spectrum.} \label{hfbkmod}
\end{center}
\end{figure}

\begin{figure*}
\begin{center}
{\epsfxsize 0.7\hsize
 \leavevmode
 \epsffile{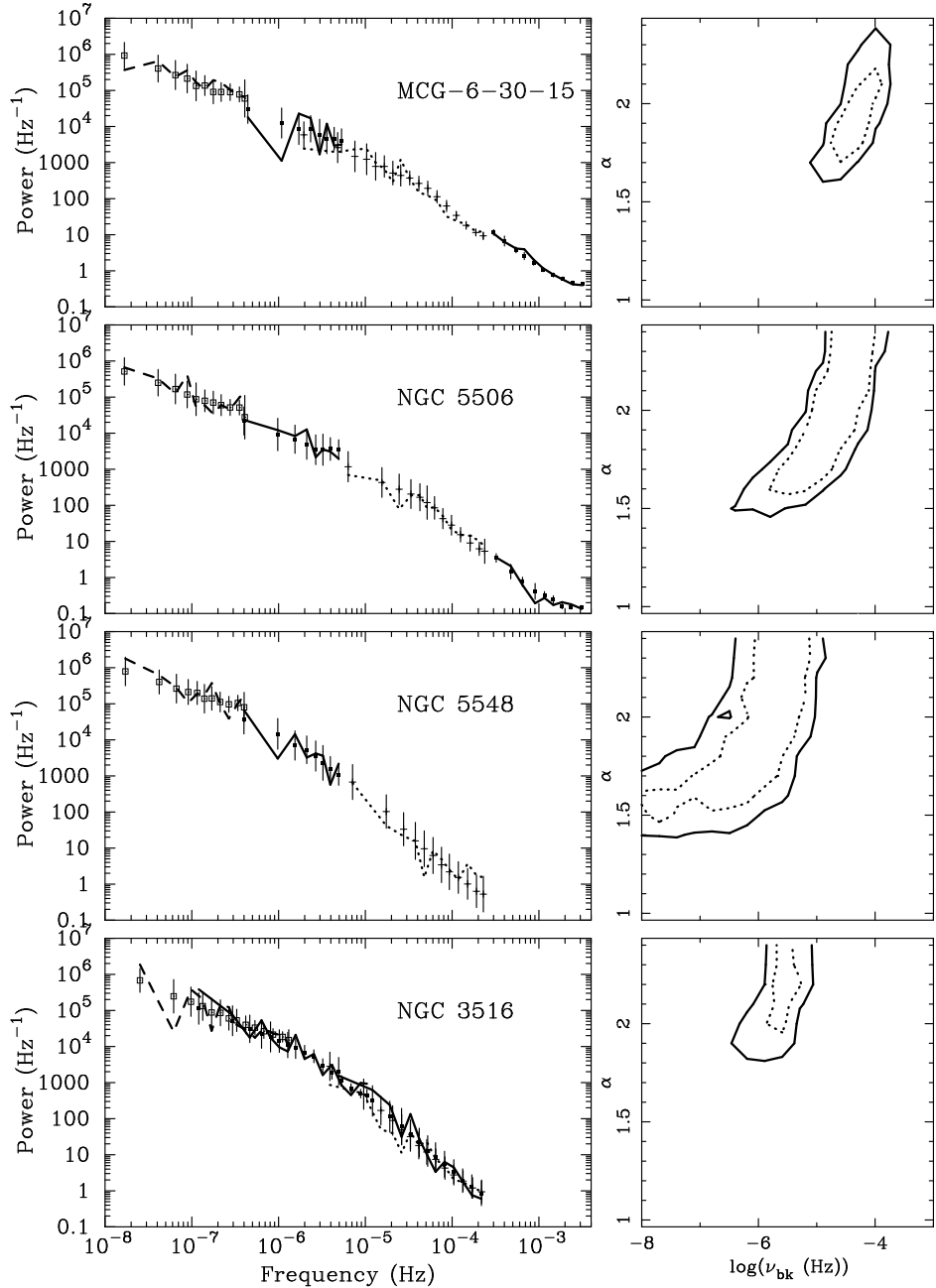}
}\caption{Comparison of best-fitting model average power spectra (points
with error bars) with 
observed power spectra (lines) for the high-frequency break
model described in the text (left),
and corresponding confidence contours for the parameter space searched 
(right).  The dashed and solid confidence contours represent the 68\% and
90\% confidence limits respectively.} \label{bkpow}
\end{center}
\end{figure*}

The motivation for the high-frequency break model comes from the power
spectrum of the black hole X-ray binary Cyg~X-1 in the low state,
which shows {\it two} frequency breaks, as described in
Section~\ref{intro}.  If AGN
have a similar power-spectral shape to Cyg~X-1 (albeit scaled down in
frequency by some factor), then because the power
spectral slopes of AGN lightcurves measured at $>10^{-5}$~Hz by
{\it EXOSAT} (e.g. Green, M$^{\rm c}$Hardy \& Lehto 1993) are
significantly greater than 1, we may be
seeing the analog of the high-frequency break in Cyg~X-1. \\
To test this possibility, we should try fitting the observed power spectra
with a model of the form used to fit the high-frequency power spectrum
of Cyg~X-1 (e.g. Nowak 1999):
\[
P(\nu)=\left\{ \begin{array}{ll}
A\,\left(\frac{\nu}{\nu_{\rm bk}}\right)^{-\alpha_{\rm hi}} & \mbox{if
$\nu>\nu_{\rm bk}$} \\
A\,\left(\frac{\nu}{\nu_{\rm bk}}\right)^{-\alpha_{\rm lo}} & 
\mbox{otherwise}
\end{array}
\right.
\]
Where $A$ is the power-spectral amplitude at the break frequency
$\nu_{\rm bk}$, and $\alpha_{\rm hi}$ and $\alpha_{\rm lo}$ are
the high and low-frequency slopes respectively, such that $\alpha_{\rm
hi}>\alpha_{\rm lo}$.  An example of a high-frequency break model with
$\alpha_{\rm hi}=2$ and $\alpha_{\rm lo}=1$ is shown in
Figure~\ref{hfbkmod}.  \\
Note that there is no physical basis for the sharpness of the
break in the high-frequency break model.  However, since the model can
adequately describe the
high-frequency power-spectral shape of Cyg~X-1, it should also serve as
a possible empirical representation of the power spectra of poorer
quality which we measure here.  We do not consider the low-frequency break in this model in order to
minimise the number of free parameters.  This approach is valid since, if the
model is correct, low-frequency breaks will occur at least a decade
lower in frequency than any measured high-frequency breaks and so will
not contribute as significantly to any flattening (besides which, if
additional low-frequency breaks are significant they will be apparent
from the residuals in any comparison of the data with the model).  \\
We only consider a low frequency slope $\alpha_{\rm lo}=1$.  Clearly
it is desirable, on
grounds of computation time, to restrict the number of free parameters
by fixing the slope below the break, but there are also compelling 
observational reasons
why we might fix the slope to $\alpha_{\rm lo}=1$.  One particularly
striking aspect of all the Cyg~X-1 power spectra is that, despite the
variations in the position of the high and low-frequency breaks and the
slope above the high-frequency break (e.g. as shown by Belloni \&
Hasinger 1990), the slope of the intermediate power-spectrum, between
the two breaks, is always remarkably close to 1.  Furthermore, Nowak et
al. (1999) show that the power spectra of Cyg X-1 made from
simultaneous lightcurves in different energy bands show an energy
dependence above the high-frequency break (in that $\alpha_{\rm hi}$
decreases towards higher energies) but maintain the same shape (i.e.
$\alpha_{\rm lo}=1$) below the break.  These results suggest that a
power-spectral slope of 1 (or very close to 1) below the high-frequency
break may, in fact, be the rule. We can
determine if the power spectra we measure are at least consistent with
this
possibility by fitting the high-frequency break model (including the
constant Poisson noise level, as before), fixing $\alpha_{\rm lo}=1$ and
stepping through the same parameter ranges as used to fit the knee model
(i.e. $\alpha_{\rm hi}=1.0$--2.4 in increments of 0.1, $\nu_{\rm
bk}=10^{-8}$--10$^{-3}$~Hz in multiples of 2). \\
The resulting best-fitting parameters are shown in
Table~\ref{bkprob}, with the results presented in the same format as for
the knee model.
\begin{table*}
\caption{Results from fitting the broadband power spectra of four Seyfert
galaxies with a high-frequency break model} \label{bkprob}
\begin{tabular}{lccc}
& Rejection confidence & $\alpha$ & $\nu_{\rm bk}/10^{-6}$~Hz \\
MCG-6-30-15 & 33\% & $2.0\pm0.3$ & 51.2 (12.8--102.4) \\
NGC~5506$^{a}$ & 10\% & $2.1\pm^{0.3}_{0.7}$ & 25.6 (0.0--102.4) \\
NGC~5506$^{b}$ & 3\% & $2.4\pm^{0}_{0.9}$ & 51.2 (0.4--102.4) \\
NGC~5548 & 27\% & $2.4\pm^{0}_{1.0}$ & 2.56 (0.0--10.24) \\
NGC~3516 & 39\% & $2.2\pm^{0.2}_{0.3}$ & 2.56 (0.64--5.12) \\
\end{tabular}

\medskip
\raggedright{See Table~\ref{kneeprob} for description.}
\end{table*}
The best-fitting model average power spectra and confidence contour
plots are shown in Figure~\ref{bkpow}.  The high-frequency break model
is an acceptable description of the data at better than 10\% confidence
for all objects in the sample.  The lower limit to break frequency is not
constrained in the power spectrum of NGC~5548.  \\

\section{Discussion}
\subsection{Summary of results}
We find that the power spectra of three of the four Seyfert galaxies in 
our sample (MCG-6-30-15, NGC~5506, NGC~3516) flatten significantly at low
frequencies and that the power-spectral flattening can be well-fitted by
either a knee or a high-frequency break model.  Although there is
no significant evidence for low-frequency flattening in the power spectrum
of NGC~5548, our model fits show that we cannot rule out the possibility
of a knee or break in the power spectrum of this object also (at
$\nu<10^{-6}$~Hz or $\nu<10^{-5}$~Hz for knee or break models
respectively). \\
We stress that the detection of low-frequency flattening
in the power spectra of AGN which we report here is completely robust,
since it is based on a rigorous Monte Carlo technique which we use to formally
reject the simple power-law model for the power spectrum.  On the other
hand, our measurements of characteristic knee or break frequencies for
the flattening are model dependent, as can be seen by the fact that two
different models for the flattening can fit the data.  Clearly, the data
are not yet adequate to distinguish between different models for the
flattening.

\subsection{Comparison with naive fits to the observed power spectrum}
\begin{figure*}
\begin{center}
{\epsfxsize 0.7\hsize
 \leavevmode
 \epsffile{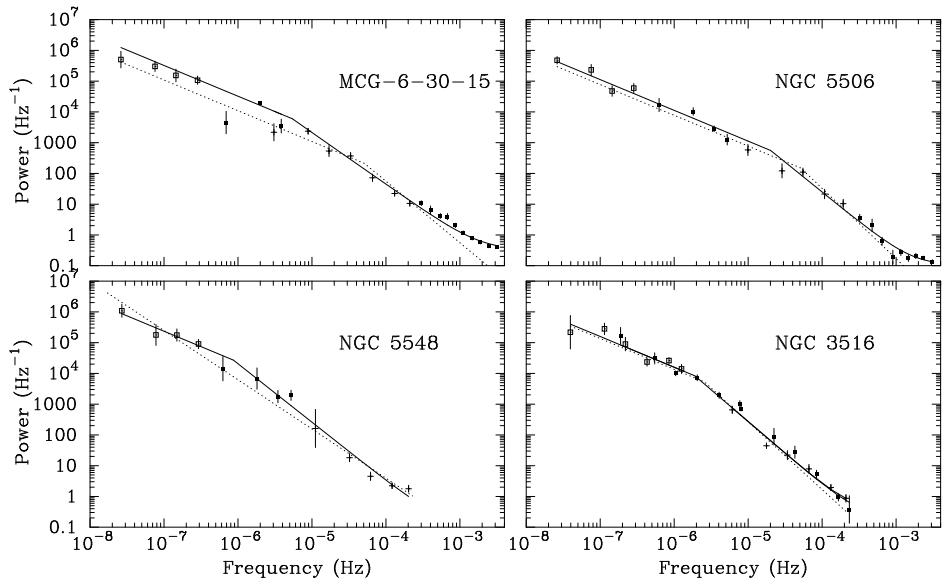}
}\caption{Best-fitting high-frequency break models (solid lines) fitted
naively to binned observed power spectra (points with error bars). 
Error bars are standard errors determined from the data.  To illustrate
aliasing effects we also show
the best-fitting underlying power spectral models obtained from the Monte Carlo
fits (dashed lines), using the broken power law model for all sources excpet
NGC~5548, where we show the simple power law
(see text for discussion).} \label{naivefit}
\end{center}
\end{figure*}

Our results confirm
earlier indications of flattening in the power spectrum of
NGC~5506 (M$^{\rm c}$Hardy 1989) and the evidence of flattening in the
power-spectra of NGC~3516 and MCG-6-30-15 presented by Edelson \& Nandra
(1999) and Nowak \& Chiang (2000) respectively.  Chiang et al. (2000) also claim
low-frequency flattening in the power spectrum of the {\it RXTE} ASM 
lightcurve of NGC~5548 but, as shown in Figure~\ref{fig:asmpca},
ASM lightcurves of
faint sources do not show the true source variability.  All these claims of
power-spectral flattening use a more
`naive' fit to the observed power spectrum, simply fitting the assumed
model directly using the data to determine errors at each frequency, and
taking no account of the distorting effects of aliasing or red-noise leak. 
The fact that we
confirm the power-spectral flattening reported in these previous works raises the
question: is it really necessary to use a Monte
Carlo technique to fit the observed power spectra? \\
To answer this
question, we naively fit the observed power spectra with the high-frequency
break model, which the {\sc
psresp} method shows is a good fit to all the data.  We bin the measured
`raw' power spectra (obtained from the rebinned and interpolated lightcurves,
see Section~\ref{powmeas}) in logarithmically spaced bins, separated by a factor of 2
in frequency (e.g. Edelson \& Nandra 1999), using a minimum of four measured
frequencies per bin.
We determine the standard error from the spread in measured powers in
each bin.  The binned power spectra and best-fitting high-frequency
break models are shown in Figure~\ref{naivefit} and comparison of the
best-fitting parameters with those obtained using Monte Carlo fits are
shown in Table~\ref{tab:naivetab}. \\

\begin{table*}
\caption{Comparison of naively fitted break model
parameters with Monte Carlo results} \label{tab:naivetab}
\begin{tabular}{lccccc}
 & \multicolumn{3}{c}{Naive results} & \multicolumn{2}{c}{Monte Carlo results} \\
& $\chi^{2}$/d.o.f. & $\alpha$ & $\nu_{\rm bk}/10^{-6}$~Hz & $\alpha$ & $\nu_{\rm bk}/10^{-6}$~Hz \\
MCG-6-30-15 & 59.6/22 & 1.68 & 5.43 & $2.0\pm0.3$ & 51.2 (12.8--102.4) \\
NGC~5506 & 25.2/20 & $1.95\pm^{0.22}_{0.16}$ & 20.0 (9.2--44.2) &
$2.4\pm^{0}_{0.9}$ & 51.2 (0.4--102.4) \\
NGC~5548 & 17.2/11 & $1.9\pm^{0.18}_{0.2}$ & 0.87 (0.18--1.9) &
$2.4\pm^{0}_{1.0}$ & 2.56 (0.0--10.24) \\
NGC~3516 & 60.16/22 & 2.06 & 1.9 & $2.2\pm^{0.2}_{0.3}$ & 2.56 (0.64--5.12) \\
\end{tabular}

\medskip
\raggedright{The table shows the $\chi^{2}$ (and degrees of freedom), slope above the
break, $\alpha$ and break frequency, $\nu_{\rm bk}$ obtained from naive model
fits, and for comparison the model parameters estimated from the Monte Carlo
fits using {\sc psresp} (see Section~\ref{hfbk}).  Errors are 90\%
confidence, corresponding to $\Delta \chi^{2}=4.61$ in the naive fits. 
Errors are not quoted where the fit is very bad.}
\end{table*}

The first point to note is that most of the error bars on the observed power
spectra are much smaller than we would expect given the spread in power
indicated by simulated power spectra.  This is because the power at each
Fourier frequency is randomly distributed with an exponential
($\chi^{2}_{2}$) distribution, hence if only a few points are sampled
the spread in points is likely to be small.  A larger number of points ($>20$)
must be averaged in each bin in order that error bars determined from
the data are reliable (see Section~\ref{errors}).  Because the errors
are underestimated, the best quality broadband power spectra
(for MCG-6-30-15 and NGC~3516, which have more extensive long-look and
intensive monitoring lightcurves respectively) are badly fitted by
the model.  Better fits are obtained for the poorer quality data (for
NGC~5506 and NGC~5548), but the 90\% confidence errors on the model
parameters are greatly underestimated, so that naive fitting implies
that the flattening in the power spectrum of NGC~5548 is significant,
whereas Monte Carlo fits show that it is not.  \\
The underestimation of errors is a major problem for naive model fitting,
which must be
taken into account when considering claims of power-spectral flattening
in the literature.  For example,
Nowak \& Chiang (2000) claim a second, significant low-frequency break at
$\sim10^{-5}$~Hz
(flattening to zero slope) in the power spectrum of MCG-6-30-15, measured
from long-look data alone.  Our simulations show that this additional
flattening is not significant.  Furthermore, our simulations show that the
best-fitting model of Nowak \& Chiang, when applied to the entire
broadband power spectrum, is ruled out at $>99$\% confidence: there is
significant power at lower frequencies than $10^{-5}$~Hz. \\
We now note the effects of aliasing on naive model fits
to observed power spectra.  To illustrate aliasing effects, we show the
best-fitting high-frequency break models from the Monte Carlo fits (i.e.
unmodified by aliasing or the Poisson noise level) in 
Figure~\ref{naivefit} as dotted lines 
(except for NGC~5548, where we show the unbroken
power law model, which was an acceptable fit to the data).  
Monte Carlo fits show that the power spectrum
of NGC~3516 is steep at high frequencies ($\alpha\sim2$).  Therefore,
since there is little high-frequency power, the amount of aliasing at lower
frequencies is small and thus the naive fitting of a break model
produces similar results to those obtained by Monte Carlo fits. 
Similarly, the naive fits to the
power spectrum of NGC~3516 carried out by Edelson \& Nandra (1999)
yield a similar knee frequency to that measured using simulations
($\nu_{\rm knee} \sim4\times10^{-7}$~Hz versus $\sim6\times10^{-7}$~Hz
respectively). \\
In contrast to the case of NGC~3516, the X-ray lightcurve of MCG-6-30-15
contains significant power up to high frequencies, and hence the effects
of aliasing are much more significant, especially in the low frequency
power spectrum,
determined from the long-term lightcurve which has the largest sampling interval. 
The result of this aliasing is a
discontinuity in the measured broadband power spectrum from the medium
to low-frequency parts of the power spectrum (reflected in the simulated
model average power spectra, see Figure~\ref{kneepow} and Figure~\ref{bkpow}),
so that the low-frequency power spectrum appears flatter and may be 
signficantly raised above the medium-frequency power spectrum.  This
effect is apparent when we compare the best-fitting model obtained from
the naive fits with that obtained by Monte Carlo fits (solid and dotted
lines respectively in Figure~\ref{naivefit}): the high frequency end of
the low-frequency power spectrum is raised significantly above the true
power level by the effects of aliasing, so that in the naive fits, 
the position of the break is pushed to significantly lower frequencies
than those estimated by the Monte Carlo fits.  The same effect might also
cause the apparent break, where none is actually required,
in the power spectrum of NGC~5548, since the power spectrum 
in this case may be unbroken and have a relatively flat slope
$\alpha=1.6$, which causes significant distortion due to aliasing in the
high-frequency ends of both the medium and low-frequency power spectra. \\
Note that the distorting effect of aliasing is made worse by the fact
that the high-frequency ends of the power spectra are also the most
well-defined, as they are made by averaging over a large number of
frequencies, so that systematic errors due to
aliasing are more pronounced than they would be if the low-frequency
ends of the power spectra were well sampled 
(since for red-noise power spectra, the aliased power at frequencies
much lower than the Nyquist frequency is small compared to
the underlying power). 
Because of this effect, particular caution should be applied to claims of
low-frequency flattening based on naive fits to the power
spectra of AGN which are highly variable on short time-scales
(e.g. Akn~564, Pounds et al. 2001).  \\
Under what circumstances might naive fitting produce reliable results
(and associated uncertainties)?  Clearly, the effects of aliasing can be
mitigated by binning up the lightcurves to a longer sampling interval,
thus smoothing out the variability on time-scales less than the sampling
interval used to make the power spectrum.  Unfortunately, the effect of
this binning is to reduce the Nyquist frequency and hence the frequency
range sampled by the power spectra.  Furthermore, since
reliable errors may only be estimated by averaging over $<20$~points per
frequency bin, the lowest frequency sampled in our observed power
spectra would have to be increased by a factor $\sim10$, in order that
sufficient low-frequency power-spectral points are averaged over. 
The net effect of binning up the lightcurves and power spectra is to
reduce the frequency range covered by the individual low and
medium-frequency power spectra by
$\sim2$~decades (i.e. to virtually a single frequency bin in each),
while the low-frequency end of the high-frequency power spectrum (which
does not suffer so much from aliasing) is shifted up by a decade. \\ 
Therefore, we conclude this discussion by noting that the naive method
may be used to fit power spectra over relatively narrow
ranges which are well
sampled (i.e. the highest frequencies in AGN long-look lightcurves),
but the Monte Carlo method of
power-spectral fitting, of which {\sc psresp} is an example, is
essential in order to place reliable constraints on the underlying
{\it broadband} power-spectral shape of AGN lightcurves obtained to date.

\subsection{Estimating black hole masses}
\label{massest}
Having measured characteristic knee or break frequencies under the assumption
of different underlying power-spectral models, we can consider the
implications of the measured characteristic frequencies for the black
hole mass.  Throughout our discussion, we make the specific assumption
that all the objects in our sample have the same power-spectral form
(knee or break) so that the characteristic frequencies are directly 
comparable. \\
What is most striking about the characteristic frequencies
measured for MCG-6-30-15 and
NGC~3516 is that they are significantly different, in either model, at a
level of better than 99\% confidence, i.e. the 90\% confidence limits of
the break frequencies do not overlap.  This is unexpected, because both
objects have similar 2--10~keV X-ray luminosities of
$L_{\rm 2-10}\sim1.5\times10^{43}$~erg~s$^{-1}$, so the fundamental 
parameter driving the position of the characteristic
frequency is not strictly related to the
luminosity.  One intriguing possibility is that although both objects
have a similar luminosity, they may have different black hole masses. 
It is quite possible that the break or knee time-scales
scale linearly with black
hole mass, as would be expected if they
correspond to a characteristic time-scale in the accretion disk or the
characteristic size scale of the system.  If this is the case, then
MCG-6-30-15 must have a significantly smaller black hole than NGC~3516
and so must be accreting at a much higher fraction of its Eddington limit. \\
Let us assume that the characteristic frequency scales linearly with the
inverse of black hole
mass, and that the mass of the black hole in Cyg~X-1 is
10~M$_{\odot}$ (Nowak et al. 1999).  We will scale the best-fitting frequency measured by the knee
model with the mean frequency of the low-frequency break in the 
low-state power spectrum of
Cyg~X-1, which varies between 0.04--0.4~Hz but has an average value of
0.13~Hz (taken from the 28 separate measurements presented in Belloni \&
Hasinger 1990).  Similarly, we scale the best-fitting frequency measured in the
high-frequency break model with the high-frequency break in the
low-state power spectrum of Cyg~X-1, with mean value 3.3~Hz, varying
between 1--6~Hz.  The resulting best estimates of black hole mass are
shown in Table~\ref{tab:masses}, together with conservative upper limits
to the black hole mass, obtained by scaling the 90\% confidence lower
limits to knee/break frequency with the respective upper values of the 
observed knee/break frequency ranges in Cyg~X-1 (i.e. 0.4~Hz and 6~Hz for
knee and break frequencies respectively). 
We also show the corresponding bolometric luminosities expressed 
as a fraction of the Eddington luminosity for a black hole with the
estimated mass, $L_{\rm
Edd}\simeq1.3\times10^{38}\,M_{\rm BH}$~erg~s$^{-1}$.
For MCG-6-30-15, we assume the bolometric luminosity of
$8\times10^{43}$~erg~s$^{-1}$, estimated from a multiwavelength study by
Reynolds et al. (1997).  For the remaining AGN, bolometric luminosities
are derived from the typical 2--10~keV X-ray
luminosities ($1.5\times10^{43}$~ergs~s$^{-1}$ for NGC~5506
and NGC~3516, $5\times10^{43}$~ergs~s$^{-1}$ for NGC~5548)
by applying the mean bolometric correction of Padovani \&
Rafanelli (1988), $L_{\rm bol}\simeq27L_{\rm 2-10}$.  Note that scatter
in this bolometric correction for individual sources, estimated from the
measurements of Padovani \& Rafanelli (1988) is possibly up to a factor
$\sim3$ in either direction.  The bolometric correction for MCG-6-30-15
is a factor 5 lower than the mean value, but Reynolds et al. (1997)
note that their estimate is a lower limit, based on the assumption of
minimal reddening in this source, so that the bolometric luminosity
in MCG-6-30-15 may be substantially larger than shown here. \\
\begin{table*}
\caption{Black hole mass and bolometric luminosity estimates} \label{tab:masses}
\begin{tabular}{lcccccccc}
& \multicolumn{4}{c}{knee model} &
\multicolumn{4}{c}{high-frequency break model} \\
& \multicolumn{2}{c}{best estimate} & \multicolumn{2}{c}{upper mass limit} &
\multicolumn{2}{c}{best estimate} & \multicolumn{2}{c}{upper mass limit} \\
& $M/10^{6}{\rm M_{\odot}}$ & $L_{\rm bol}/L_{\rm Edd}$ & 
$M/10^{6}{\rm M_{\odot}}$ & $L_{bol}/L_{\rm Edd}$ &
$M/10^{6}{\rm M_{\odot}}$ & $L_{\rm bol}/L_{\rm Edd}$ &
$M/10^{6}{\rm M_{\odot}}$ & $L_{\rm bol}/L_{\rm Edd}$ \\
MCG-6-30-15 & 0.25 & 2.4 & 1.6 & 0.4 & 0.6 & 1 & 5 & 0.14 \\ 
NGC~5506 & 0.5 & 6 & 25 & 0.1 & 0.6 & 5 & 150 & 0.02 \\
NGC~5548 & 65 & 0.2 & NA & NA & 13 & 0.8 & NA & NA \\
NGC~3516 & 2 & 1.5 & 13 & 0.25 & 13 & 0.25 & 94 & 0.03 \\
\end{tabular}

\medskip
\raggedright{See text for details.  Mass estimates for NGC~5506 use the model fit
results for the {\it RXTE} plus {\it EXOSAT} data.  No upper limits can
be set on the black hole mass of NGC~5548.}
\end{table*}
Table~\ref{tab:masses} shows that, for the objects with the
best-constrained characteristic frequencies in their power spectra
(MCG-6-30-15 and NGC~3516), the knee model predicts a lower black hole
mass than the high-frequency break model.  Reverberation mapping
estimates of the mass of NGC~3516 suggest a black hole mass of
$\sim2\times10^{7}$~M$_\odot$ (Wanders \& Horne 1994), compatible with
the best estimate of mass from the high-frequency break model, but not
inconsistent with the upper limit derived from the knee model.  Given the
uncertainty in the true bolometric
luminosity, and the conservative upper mass limits quoted here, the
masses predicted by the knee model are consistent with sub-Eddington
accretion rates. \\
However, the conservative 40\% Eddington lower limit to accretion
rate estimated from the knee model fits to the power spectrum of 
MCG-6-30-15 (where the bolometric luminosity quoted represents a
firm lower bound) is in contrast to that observed in
Cyg~X-1 where the transition between the low and high states occurs
at only a few per cent of the
Eddington luminosity (Di Salvo et al. 2001, Zhang et al. 1997).    
Since the knee model represents the low-frequency break
seen in the power spectrum of
Cyg X-1 in the low state, it would seem that the evidence from accretion
rate is inconsistent with the knee model, at least in this source.
Interestingly, even the high-frequency break model
predicts an upper limit to black hole mass which is only barely consistent with 
the low-state accretion rates seen in Cyg~X-1.  
An intriguing possibility is that MCG-6-30-15 is in a state
analogous to the high state of Cyg~X-1, which has a characteristic steep
($\alpha\simeq2$) high-frequency power spectrum breaking at 10~Hz 
to $\alpha\simeq1$, with no low-frequency breaks down to frequencies as
low as $10^{-2}$~Hz (Cui et al. 1997).  Therefore, high state power spectra
similar in shape to that of Cyg~X-1 would also be compatible with the
high-frequency break model we use here.  Scaling
by the high state frequency-break yields an estimated black hole mass of
$2\times10^{6}$~M$_{\odot}$ (upper limit
$\sim8\times10^{6}$~M$_{\odot}$), implying an accretion rate of 
$\sim30$\% Eddington (which we might expect from a high-state source).  \\
Although the black hole mass in MCG-6-30-15
has not yet been measured by reverberation mapping, Reynolds (2000) has
pointed out that the low luminosity (absolute B magnitude $\simeq-19$)
of the S0 host galaxy of this AGN
corresponds to a black hole mass of $\sim10^{7}$~M$_{\odot}$ if the
black hole mass-bulge luminosity relation is the same as that determined by
Magorrian et al. (1998) for a sample of normal elliptical and S0 galaxies.  We
note here that the black hole masses of all the Seyfert galaxies which
have been reverberation mapped show smaller black hole masses by a
factor of $\sim10$ than the Magorrian relation suggests (Wandel 1999). 
This discrepancy would imply a black hole mass for MCG-6-30-15 which is
significantly lower than $10^{7}$~M$_{\odot}$, consistent with the
low mass suggested by the high state interpretation of the power
spectrum.  We further note that the lower, reverberation mapping estimates
of black hole mass in AGN also follow the recently discovered correlation
of black hole mass with bulge velocity dispersion, which has been found
to be a better predictor of black hole mass than the Magorrian relation
(Gebhardt et al. 2000, Ferrarese et al. 2001).  Therefore it seems
plausible that the black hole in MCG-6-30-15 has a mass of order
$10^{6}$~M$_{\odot}$.  Reverberation mapping, or another independent mass
estimation technique will be required to confirm this.   \\
The power-spectral shape of NGC~5506 is also consistent with a high
state interpretation of this source, although the uncertainties on the
break frequency are sufficiently large that a low-state interpretation
would suffice to fit the data self-consistently.  We note here however
that the low luminosity of the NGC~5506 galaxy (absolute B magnitude
$\simeq-20$), suggests that the central black hole mass is relatively
low.  Although reverberation mapping of NGC~5506 is not possible because
it is a Seyfert~2, it does have a water megamaser (Braatz,
Wilson \& Henkel 1994) which might permit a good estimate of its 
central black hole mass for comparison with the mass estimated from its
power spectrum.  \\
The high state AGN interpretation for either
MCG-6-30-15 or NGC~5506 is highly speculative at this stage.  One
significant problem is that the energy spectra of neither source
shows the steep power-law 
($\Gamma\sim2.6$) and strong disk blackbody components seen in the
spectrum of Cyg~X-1 in
the high state (Cui et al. 1998), although this difference might be
a result of the significantly lower disk temperature expected in 
AGN compared to BHXRBs.  Greater
support for the high state interpretation in either MCG-6-30-15 and
NGC~5506 could be provided by the detection of 
significant power down to
very low frequencies, implying that there is no low-frequency
break in the power spectrum on frequencies $<0.01\nu_{\rm bk}$. 
Continued long-term monitoring will help to resolve this issue.    \\
The frequency break in NGC~5548 is not signficantly detected.  
The black hole mass of NGC~5548 measured from
reverberation mapping is $\sim10^{8}$~~M$_{\odot}$ (Wandel, Peterson \&
Malkan 1999), implying that the accretion rate is
relatively low ($<10\%$ Eddington), so applying the low state
interpretation we can estimate that the high-frequency break occurs at
$\sim2\times10^{-7}$~Hz, and may become detectable with further monitoring
observations. \\

\subsection{The AGN-BHXRB connection and physical implications}
The main result of this work has been to show that the power spectra of
AGN flatten towards low frequencies.  Whether this flattening has the
same form as that in BHXRBs is still not certain.  The data do not yet
allow us to categorically rule out either knee or high-frequency break 
models, let alone more complex
models where the power spectra might have multiple breaks or flatten
gradually.  However, it is encouraging that a simple high-frequency
break, similar to that seen in the classic BHXRB Cyg~X-1,
can fit the data well, and yield plausible black hole masses if we assume that
the break frequency scales inversely with black hole mass.  \\
\begin{figure*}
\begin{center}
{\epsfxsize 0.9\hsize
 \leavevmode
 \epsffile{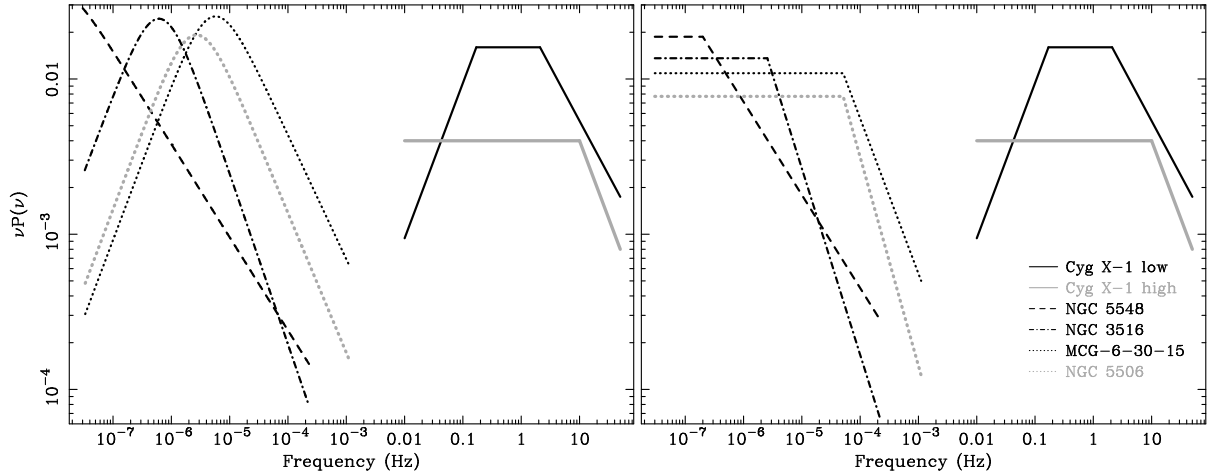}
}\caption{Comparison of best-fitting knee (left panel) 
and high-frequency break (right panel) model
$\nu P(\nu)$ power spectra of MCG-6-30-15,
NGC~5506, NGC~3516, NGC~5548 with power spectra of Cyg~X-1 in the low and high
states.}  \label{nfncomp} 
\end{center}
\end{figure*}
A useful way of comparing power spectra is to plot
frequency$\times$power, rather than power, as a function of frequency. 
The $\nu P(\nu)$ plot produced in this way is analogous to the $\nu F(\nu)$
method of displaying energy spectra, in that the peak in the $\nu P(\nu)$
plot shows which time-scales most of the variability occurs on, as well
as the magnitude of variability on those time-scales.  In
Figure~\ref{nfncomp} we plot the $\nu P(\nu)$ power spectra for
the best-fitting knee and high-frequency break models fitted to the observed
power spectra of MCG-6-30-15, NGC~5506 and NGC~3516
(see Section~\ref{hfbk}).  The $\nu P(\nu)$ power spectra of NGC~5548 are also
included in the figure, except that since the knee or break is not
defined for this object, we have assumed knee and break frequencies of
$10^{-8}$~Hz and $2\times10^{-7}$~Hz respectively,
corresponding to the black hole mass of $10^{8}$~M$_{\odot}$ estimated
from reverberation mapping.  For comparison, $\nu P(\nu)$ power
spectra of Cyg~X-1
are included, corresponding to typical power-spectral parameters in the
low and high states (Nowak et al. 1999, Cui et al. 1997). \\
Note that the flat peaks in the high-frequency break $\nu P(\nu)$
power spectra of AGN and the typical power spectra of Cyg~X-1 correspond to
the $\alpha=1$ part of the power spectrum, where there is equal
integrated power per decade of frequency.  The low-frequency 
drop-off in the $\nu P(\nu)$ power in the low state of Cyg~X-1
corresponds to the low-frequency break or `knee' in the power spectrum,
which we have not yet detected in the AGN power spectra if we are
seeing the high-frequency break, because our low-frequency data is
not yet adequate to detect it.  Therefore, until clear evidence for 
low-frequency flattening to a slope $\alpha=1$ is found, we cannot rule out the
possibility that all our sources have power spectra similar to the high
state power spectrum in Cyg~X-1. \\
Note that the knee model $\nu P(\nu)$ power spectra do not show flat
peaks, because the high-frequency slopes fitted are all significantly 
steeper than 1.  If we were measuring the low-frequency break alone
(i.e. the frequency of the high-frequency break is too high to detect), we
would expect the knee model to fit high-frequency slopes of $\alpha\simeq1$.
This result provides additional evidence that we are 
indeed measuring the high-frequency breaks in these sources, if
the power spectra of AGN have the same shape as those of BHXRBs. However,
we stress that we cannot yet rule out the possibility that we are
measuring more complex configurations including both high and
low-frequency breaks. \\
Fig.~\ref{nfncomp} shows clearly that the $\nu P(\nu)$
power spectra of AGN are similar to those in Cyg~X-1, in that they
have similar peak powers, but are shifted at least 5 decades
down the frequency axis.  The fact that the peak powers are similar
implies that the number of varying regions and the general pattern of
variability is the same in AGN and BHXRBs, while the luminosity of the
varying regions and their variability time-scales are scaled by some
factor which could be the black hole mass.

\section{Conclusions}
We have presented long-term {\it RXTE} monitoring lightcurves for 4
Seyfert galaxies, MCG-6-30-15, NGC~5506, NGC~5548 and NGC~3516, and
measured their broadband power spectra to determine if they flatten
towards low frequencies, like those of BHXRBs.  The interpretation of
power spectra measured from discretely (and sometimes unevenly) sampled
lightcurves is complicated by the distorting effects of red-noise leak
and aliasing.  A further serious problem is that reliable errors in the power
in each frequency bin cannot be defined from the data, due to the small
number ($<20$) of power-spectral measurements in each frequency bin.  \\
To overcome these difficulties, we have built on the response method of
Done et al. (1992), to develop a reliable Monte Carlo method for testing
models for the underlying power-spectral shape of discretely, unevenly 
sampled lightcurves.  Our method, {\sc psresp}, takes proper account of
the effects of aliasing and red-noise leak and crucially, uses the
distribution of simulated power spectra to define reliable confidence
limits on our model fits. \\
We have used {\sc psresp} to test simple models for the power-spectral
shape of the active galaxies in our sample.  Our main results are:
\begin{enumerate}
\item[1.] A single power-law model for the power spectra, with no
low-frequency flattening, is rejected at
better than 95\% confidence for MCG-6-30-15, NGC~5506 and NGC~3516.  The
power spectrum of NGC~5548 is consistent with a single power-law.
\item[2.] Both the knee model (flattening to slope $\alpha=0$) and high-frequency break
model (flattening to $\alpha=1$) provide good fits
to the power spectra of all four sources.  Knee/break frequencies are
well constrained for all sources except NGC~5548, for which we can define
upper limits only.
\item[3.] The characteristic knee/break frequency measured for MCG-6-30-15 is
significantly higher (99\% confidence) than the corresponding frequency in NGC~3516, even
though both sources have similar X-ray luminosities ($\sim1.5\times
10^{43}$~erg~s$^{-1}$).
\item[4.] If the knees or breaks correspond to those seen in the
low-state power spectrum of Cyg~X-1, and the characteristic knee/break
time-scales scale linearly with black
hole mass, the black hole masses estimated from the knee model are lower
than those estimated by the break model, but remain consistent with
sub-Eddington accretion rates.  In the case of MCG-6-30-15, the
conservative lower limit on accretion rate estimated from the knee model
is an order of magnitude higher than that seen in the low state of
Cyg~X-1, favouring the high-frequency break model and further suggesting 
that the break seen in
the power spectrum of MCG-6-30-15 (and possibly NGC~5506) is analogous to
the high-frequency break seen in the {\it high state} power spectrum of
Cyg~X-1. 
\item[5.] The $\nu P(\nu)$ power spectra of the Seyfert galaxies studied
here are similar in amplitude to the $\nu P(\nu)$ power spectra of
Cyg~X-1 in the high and low states.
\end{enumerate}
Conclusions 4 \& 5 imply that the power spectra of AGN are consistent
with being identical in shape and fractional RMS amplitude (integrated
over the whole power spectrum) to those of BHXRBs, with
characteristic time-scales scaling linearly with black hole mass. 
Arguments based on accretion rate seem to favour the high-frequency break
model over the knee model in at least one source (MCG-6-30-15), although
this evidence is circumstantial.  Further monitoring observations (which are
currently underway) are needed to
reject either model on the basis of power-spectral measurements
alone, and to confirm the interesting possibility that two of the
objects in our sample are analogous to high-state BHXRBs. 

\subsection*{Acknowledgments}
We would like to thank the entire {\it RXTE} team for making this work
possible, and the anonymous referee, for helpful comments and
suggestions.  This research has made use of data obtained from the High
Energy Astrophysics Science Archive Research Center (HEASARC), provided by
NASA's Goddard Space Flight Center.

\bsp
\end{document}